\documentclass[5p,twocolumn]{elsarticle}
\usepackage{soul}

\usepackage{indentfirst}
\usepackage{amsmath}
\usepackage{amsfonts}
\usepackage{amssymb}
\usepackage{graphicx}
\usepackage{hyperref}
\usepackage{mathtools}
\usepackage{verbatim}
\usepackage{epstopdf}
\usepackage{subfigure}
\usepackage{latexsym}
\usepackage{dcolumn}
\usepackage{epsf}
\usepackage{float}
\usepackage[table]{xcolor}
\usepackage{multirow}
\usepackage{makecell}
\usepackage{color}
\usepackage{bm}
\usepackage{rotating}

\usepackage{multicol}

\usepackage[normalem]{ulem}
\usepackage{xcolor}
\definecolor{forestgreen}{rgb}{0.33,0.61,0.34}

\usepackage[normalem]{ulem}
\usepackage{xcolor} 

\usepackage{url}
\usepackage[utf8]{inputenc}
\biboptions{numbers,sort&compress}  
\usepackage{times}
\usepackage{sectsty}
\usepackage{stackengine}
\usepackage{caption}
\allsectionsfont{\normalsize} 
\makeatletter
\def\ps@pprintTitle{%
 \let\@oddhead\@empty
 \let\@evenhead\@empty
 \def\@oddfoot{}%
 \let\@evenfoot\@oddfoot}

\makeatother

\begin{document}
\begin{frontmatter}

\title{State-transition dynamics of resting-state functional magnetic resonance imaging data: Model comparison and test-to-retest analysis}
\author{%
Saiful Islam$^{2}$, \ Pitambar Khanra$^1$, \ Johan Nakuci$^{4}$, \ Sarah F. Muldoon$^{1,2,3}$,  \ Takamitsu Watanabe$^5$,\ Naoki Masuda$^{1,2*}$
}

\address{
$^1$Department of Mathematics, University at Buffalo, State University of New York at Buffalo, USA\\
$^{2}$Institute for Artificial Intelligence and Data Science, University at Buffalo, State University of New York at Buffalo, USA\\
$^3$Neuroscience Program, University at Buffalo, State University of New York at Buffalo, USA\\
$^4$School of Psychology, Georgia Institute of Technology, USA\\
$^5$International Research Centre for Neurointelligence, The University of Tokyo Institutes for Advanced Study, Tokyo, Japan\\
$^*$Corresponding author: naokimas@buffalo.edu  
}

\begin{abstract}
Electroencephalogram (EEG) microstate analysis entails finding dynamics of quasi-stable and generally recurrent discrete states in multichannel EEG time series data and relating properties of the estimated state-transition dynamics to observables such as cognition and behavior. While microstate analysis has been widely employed to analyze EEG data, its use remains less prevalent in functional magnetic resonance imaging (fMRI) data, largely due to the slower timescale of  such data. In the present study, we extend various data clustering methods used in EEG microstate analysis to resting-state fMRI data from healthy humans to extract their state-transition dynamics. We show that the quality of clustering is on par with that for various microstate analyses of EEG data. We then develop a method for examining test-retest reliability of the discrete-state transition dynamics between fMRI sessions and show that the within-participant test-retest reliability is higher than between-participant test-retest reliability for different indices of state-transition dynamics, different networks, and different data sets. This result suggests that state-transition dynamics analysis of fMRI data could discriminate between different individuals and is a promising tool for performing fingerprinting analysis of individuals. 
\end{abstract}


\end{frontmatter}

\section{Introduction}\label{sec:introduction}
Activity of the human brain is dynamic even at rest, and brain dynamics on various spatial scales are considered to drive myriad functions of the brain \cite{raichle2001default, rabinovich2006dynamical, deco2015rethinking, avena2018communication}. Multiple methods to characterize brain dynamics have been proposed, many of which rely on the detection of brain states and quantification of how the brain transitions through such states. Microstate analysis is an early-proposed method for estimating discrete states in electroencephalogram (EEG) data~\cite{khanna2015microstates, michel2018eeg, von2018eeg}. EEG microstate analysis usually entails clustering of multi-electrode EEG signals, with each data point to be clustered corresponding to a time point of the measurement. Each cluster, or microstate, is a representation of a global functional state of the brain. Microstates obtained from resting-state EEG data tend to last about 100 ms and are reproducible \cite{koenig2002millisecond, khanna2014reliability, michel2018eeg, liu2020reliability, zhang2021reliability}. Microstate analysis has been extended for magnetoencephalography (MEG) data, with the microstates being estimated by conventional clustering methods \cite{tait2022tool, tait2022meg} or the hidden-Markov model (HMM)~\cite{baker2014fast, tait2022tool} among other methods.  
Microstate analysis in its original sense (i.e., detecting and utilizing microstates lasting about 100 ms) does not directly apply to functional magnetic resonance imaging (fMRI) data because the temporal resolution of fMRI is limited, preventing one from detecting dynamics on the timescale of 100 ms. One direction to resolve this limitation is to use EEG microstate analysis results to inform states in fMRI data \cite{britz2010bold, schwab2015discovering,  case2017characterization, brechet2019capturing}. 
An alternative approach is to estimate and use state-transition dynamics of spatial fMRI signals, as microstate analysis does for EEG (and MEG) data, regardless of different time resolutions between fMRI and EEG/MEG. Such state-transition dynamics for fMRI data have been estimated by data clustering algorithms as in the case of the EEG/MEG microstate analysis \cite{allen2014tracking, calhoun2014chronnectome,  nielsen2018predictive, ezaki2021modelling}, the HMM or its variants \cite{ryali2016temporal, vidaurre2017brain, taghia2017bayesian, brookes2018altered, nielsen2018predictive, vidaurre2021new, ezaki2021modelling}, and energy landscape analysis \cite{watanabe2013energy, watanabe2014energy, ezaki2017energy, ezaki2018ge}. Each discrete state in fMRI data corresponds to a vector of activity patterns at specified regions of interests (ROIs) \cite{liu2013decomposition, liu2013time,  liu2018co, ezaki2021modelling, paakki2021co}, or a functional network among ROIs~\cite{sakouglu2010method, hutchison2013resting, leonardi2013principal, calhoun2014chronnectome, allen2014tracking, abrol2016chronnectome, ryali2016temporal, taghia2017bayesian, nielsen2018predictive}.

In general, successful single individual inferences from neuroimaging data would suggest their potential applications for both scientific investigations and clinical practice. Research has shown that functional networks from fMRI data can be used as a reliable fingerprint of human individuals through test-retest analyses \cite{finn2015functional,  gordon2017precision, amico2018quest, noble2019decade, venkatesh2020comparing, chiem2022improving}. Test-retest reliability has also been assessed for dynamic functional networks estimated from fMRI data~\cite{zhang2018test, zhang2022test, long2023evaluating}, whereas test-retest reliability for dynamic functional networks has been reported to be lower than that for static functional networks~\cite{zhang2018test, zhang2022test}. With this study, we are interested in test-retest reliability of state-transition dynamics in fMRI data, which has been underexplored.

In the present study, we assess the potential effectiveness of dynamics of discrete states estimated from fMRI data at fingerprinting individuals. Here, we use fMRI data as multivariate time series, each dimension of which represents a single ROI, akin to microstate analysis for EEG and MEG data. This approach contrasts with the aforementioned prior studies on test-retest reliability of dynamic functional networks. Our analysis involves examination of what methodological choices (e.g., the clustering method applied to the fMRI data to define discrete states, the number of clusters identified, and the indices used to characterize the estimated state transition dynamics) yield a higher test-retest reliability of the state-transition dynamics; such an assessment has previously been carried out for EEG microstate analysis \cite{khanna2014reliability}. Based on a permutation test to quantify test-retest reliability, we show that, in general, transitory dynamics of discrete states estimated for fMRI data yield higher within-participant than between-participant test-retest reliability across clustering methods, the number of clusters, observables of the state-transition dynamics, two sets of ROIs, and two data sets. Code for computing dynamics of discrete states and their test-retest reliability used in the present paper is available on GitHub ~\cite{SLIslam2023}.

\section{Methods}
\label{sec:methods}
\subsection{Midnight Scan Club data} 
\label{sub:MSC}
We use the resting-state fMRI data provided by the Midnight Scan Club (MSC) project~\cite{gordon2017precision}. The MSC's resting-state fMRI data consist of recording from ten healthy human adults (age: $29.1\pm3.3$ (average ± standard deviation); five males and five females) over ten consecutive nights. A single recording session of the resting-state fMRI experiment lasted for 30 mins, resulting in 818 volumes. The imaging was performed on a Siemens TRIO 3T MRI scanner. All functional imaging was performed using an echo planar imaging (EPI) sequence (TR $= 2.2$ s, TE $= 27$ ms, flip angle $= 90^{\circ}$, voxel size $= 4$ mm $\times$ 4 mm $\times$ 4 mm, 36 slices).

It was originally reported that the eighth participant (i.e., MSC08) fell asleep, showed frequent and prolonged eye closures, and had systematically large head motion, yielding considerably less reliable data than those obtained from the other participants~\cite{gordon2017precision}. In our previous work, we also noticed that the quality of the data analysis fluctuated considerably more across the different sessions for the tenth participant (i.e., MSC10) than for the other participants except MSC08~\cite{khanra2023reliability}. Therefore, we excluded MSC08 and MSC10  and only used the remaining eight participants (age: $29.1\pm3.4$; four males and four females) in the following analysis.

We used SPM12 (http://www.fil.ion.ucl.ac.uk/spm) to preprocess the resting-state functional images. Specifically,
we first conducted realignment, unwrapping, slice-timing correction, and normalization to a standard template (ICBM 152). Then, we performed regression analyses to remove the effects of head motion, white matter signals, and cerebrospinal fluid signals. Lastly, we conducted band-pass temporal filtering (0.01--0.1 Hz).

We used a DMN composed of 12 ROIs~\cite{fair2009functional}. To optionally reduce the dimension of the DMN, which may improve the estimation of discrete states, we averaged over each pair of the symmetrically located right- and left-hemisphere ROIs into one observable. The symmetrized DMN has eight ROIs because four ROIs (i.e., amPFC, vmPFC, pCC, and retro splen) in the original coordinate system are approximately on the midline and therefore have not undergone the averaging over the right- and left-hemisphere ROIs~\cite{ezaki2018ge}.  

In addition to the DMN, we also analyzed the so-called whole-brain network. We determined the regions of interest (ROIs) of the whole-brain network by employing the 264 spherical ROIs whose coordinates were identified in a previous study~\cite{power2011functional}. We then removed 50 ROIs labelled `uncertain' or `subcortical', resulting in 214 ROIs. We removed these 50 ROIs because they are located in the peripheral or border areas of the brain such that signals from these regions are easily contaminated with cerebrospinal fluid (CSF) signals.
The 214 ROIs were labeled either of the following nine functionally different brain systems:
auditory network, dorsal attention network (DAN), ventral attention network (VAN), cingulo-opercular network (CON), default mode network (DMN), fronto-parietal network (FPN), salience network (SAN), somatosensory and motor network (SMN), or visual network. We merged the DAN, VAN, and CON into an attention network (ATN) to reduce the number of observables from nine to seven, as we did in our previous studies~\cite{watanabe2017brain,khanra2023reliability}.
This is because the DAN, VAN, and CON have been suggested to be responsible for similar attention-related cognitive activity~\cite{power2011functional}. We then calculated the average fMRI signal for each of the seven systems by first averaging the signal over the volumes in the sphere of radius 4 mm centered around the provided coordinate of each ROI~\cite{power2011functional}, and then averaging the signal over all ROIs belonging to the system (e.g., 13 ROIs in the auditory network). We call the thus obtained seven-dimensional system the whole-brain network. It should be noted that the DMN constitutes one ROI in the whole-brain network, whereas the DMN described above as a system of ROIs is composed of either 8 or 12 ROIs depending on whether or not we average over symmetrically located ROIs.

\subsection{Human Connectome Project data} \label{sub:HCP}

We also analyzed the fMRI data recorded from healthy human adults and shared as the S1200 data in the Human Connectome Project (HCP)~\cite{van2013wu}. In the S1200 data, 1200 adults between $22$–$35$ years old underwent four sessions of 15-min EPI sequence with a 3T Siemens Connectome-Skyra (TR $= 0.72$ s, TE $= 33.1$ ms, $72$ slices, $2.0$ mm isotropic, field of view (FOV) $= 208 \times 180$ mm) and a T1-weighted sequence (TR $= 2.4$ s, TE $= 2.14$ ms, $0.7$ mm isotropic, FOV $= 224 \times 224$ mm). To circumvent any potential influence of familial relationships, we use a subset of 100 participants that do not share any family relationship, called ``100 unrelated participants'' (age: $29.4\pm3.5$; 46 males and 54 females), released by the HCP. All these 100 participants completed both diffusion weighted MRI and two resting-state fMRI scans.

Each participant underwent two sessions of resting-state fMRI recording, and each session consisted of both Left-Right (LR) and Right-Left (RL) phases.  In the following text, we refer to phases as sessions. Therefore, each participant's data consist of four sessions. We used data from participants with at least 1150 volumes in each of the four sessions after we had removed volumes with motion artifacts,  resulting in a final analysis of 87 participants (age: $29.8\pm3.4$; 40 males and 47 females). For the 87 participants, we removed the volumes with motion artifacts and then used the last 1150 volumes in each session, with the aim of removing possible transient effects.

We employed independent component analysis (ICA) to remove nuisance and motion signals~\cite{glasser2013minimal}. Then, any volumes with frame displacement greater than 0.2 mm~\cite{jenkinson2002improved} were excised~\cite{power2012spurious}. This is because the ICA-FIX pipeline has been found not to fully remove motion-related artifacts~\cite{burgess2016evaluation, siegel2017data}. Next, we standardized each voxel by subtracting the temporal mean, and then global signal regression (see section~\ref{sub:gsr}) was carried out.

We averaged the fMRI signal over all the voxels within each ROI of the AAL atlas~\cite{tzourio2002automated} in each volume. We remark that the AAL atlas is composed of 116 ROIs. In order to map these ROIs to representative brain systems, we first mapped each of the cortical ROIs to the parcellation scheme from the Schaefer-100 atlas~\cite{schaefer2018local}. We based the assignment of the ROI to the brain system that minimized the Euclidian distance from the centroid of an ROI in the AAL to the corresponding centroid of an ROI in the Schaefer atlas. We provide the correspondence between each AAL ROI and the brain system in Supplementary Information (see section S1). After we assigned each ROI to a system, we removed 42 ROIs labeled `subcortical' or `cerebellar', which yielded 74 ROIs. These 74 ROIs were then assigned to one of the $N=7$ functionally different brain networks: control network, DMN, DAN, limbic network, salience/ventral attention network, somatomotor network, and visual network. We call this seven-dimensional system the whole-brain network for the HCP data. Similarly to the case of the whole-brain network for the MSC data, we first averaged the fMRI signal over the voxels within each ROI and then further averaged the signal over the ROIs belonging to the same system (e.g., 59 ROIs belonging to the DMN). 

\subsection{Global signal removal} \label{sub:gsr}

We denote the fMRI time series for a session by $\{ \mathbf{x}_1, \ldots, \mathbf{x}_T \}$, where $T$ is the number of volumes (i.e., time points), $\mathbf{x}_t = (x_{t,1}, \ldots, x_{t,\tilde{N}})$ is the fMRI signal at time $t$, and $\tilde{N}$ is the number of ROIs with which we compute the global signal. Note that $\tilde{N}$ may be larger than $N$, which occurs when we define a global signal widely from the brain including ROIs that we do not use for estimating the discrete states.
The global signal is the average of the signal over all the $\tilde{N}$ ROIs at each time, i.e.,
\begin{equation} \label{eqn:gs}
\overline{x}_t = \frac{\sum_{i=1}^{\tilde{N}} x_{t, i}}{\tilde{N}}.
\end{equation}

We remove the global signal~\cite{murphy2017towards} by subtracting $\overline{x}_t$ from each $x_{t, i}$ (with $i \in \{1, \ldots, \tilde{N} \}$) and dividing the result by
the standard deviation, i.e.,  
\begin{equation} \label{eqn:gfp}
\sigma_t = \sqrt{\frac{ \sum_{i=1}^{\tilde{N}} (x_{t,i} -\overline{x}_t)^2} {\tilde{N}}}.
\end{equation}
We carry out this procedure for each $t$.

The global signal in resting-state fMRI data is considered to primarily consist of physiological noise stemming from different factors such as respiration, scanner-related artifacts, and motion-related artifacts. By removing the global signal, several quality-control metrics are improved, the anatomical specificity of functional-connectivity patterns is enhanced, and there is a potential increase in behavioral variance~\cite{li2019topography,aquino2020identifying}.

It is a common practice to calculate the global signal using the gray matter, white matter, and CSF (e.g., \cite{fair2009functional, gordon2017precision}). However, we calculate the global signal using the gray matter but not the white matter or CSF~\cite{siegel2017data}, as explained above, for the following reasons. First, white matter noise and CSF-oriented noise were removed in the preprocessing procedure that aims to suppress the effect of motion. Second, white matter and CSF signals contribute little to the fMRI signal in the gray matter. In fact, the characteristics of the blood-oxygen-level-dependent (BOLD) signal are substantially different in the gray matter compared with the white matter and even more so with CSF, which should not contain any signal associated with the BOLD signal. Therefore, including the white matter and CSF in the global signal calculation is more likely to induce distortions in the signal~\cite{glasser2013minimal, power2014methods}.

For the DMN obtained from the MSC data, we first removed the global signal calculated over the $\tilde{N} = 30$ ROIs in the coordinate system provided by~\cite{fair2009functional}, which included the $N=12$ ROIs in the DMN. Then, we compared three treatments of global signal removal for the DMN as follows. In the first and second treatments, we then removed the global signal calculated from the $\tilde{N} = 12$ DMN ROIs from each of the $12$ ROIs in the DMN. Next, we averaged the obtained time series over each symmetric pair of DMN ROIs corresponding to the two hemispheres. If the ROI is roughly on the midline, there is no such symmetric pair of ROIs, in which case we only removed the global signal. After aggregating the symmetric pairs of ROIs in this manner, there are $N=8$ ROIs in the DMN. This concludes the first treatment. In the case of the second treatment, we additionally removed the global signal calculated over the $\tilde{N} = N = 8$ ROIs. In the third treatment, after the removal of the global signal calculated over $30$ ROIs, which is common for all the three treatments, we further removed global signal calculated from the $\tilde{N}=12$ DMN ROIs from each of the 12 ROIs. We do not further process the data. Therefore, with the third treatment, the final DMN consists of $N=12$ ROIs.

For the whole-brain network obtained from the MSC data, we first removed the global signal computed from the $\tilde{N} = 264$ ROIs. Then, we extracted $N=7$ -dimensional time series as described in section~\ref{sub:MSC}. Finally, we further removed the global signal computed from the $\tilde{N} = N = 7$ ROIs in the whole-brain network. The global signal removal for the whole-brain network obtained from the HCP data is the same except that we computed the first global signal from
the $\tilde{N} = 116$ ROIs of the AAL atlas (see section~\ref{sub:HCP}).

\subsection{Estimation of discrete states}\label{subsec:micro analy}

There are various methods for estimating microstates in the EEG and MEG data \cite{khanna2015microstates, von2018eeg, michel2018eeg, tait2022meg}. We tailor seven popular methods for finding microstates in EEG and MEG data to the case of fMRI data to estimate their discrete states. Because the discrete states that we find for fMRI data are not equivalent to EEG/MEG microstates, we refer to the former as states, discrete states, or clusters in the following text. We describe each method in the following subsections. See Table~\ref{tab:notation} for main notations.

\begin{table}
\centering
\caption{Main notations used in this paper.}
\setlength{\tabcolsep}{1pt}
\begin{tabular}{ |c|>{\centering\arraybackslash}m{7.5cm}| } 
 \hline
Symbols  & Description \\ 
\hline
$N_\text{p}$& Number of participants \\
$N_\text{s}$& Number of sessions for each participant \\
$N$& Number of ROIs \\
$T$& Number of volumes (i.e., time points) in each session \\
$\bm{x}_{t}  \in \mathbb{R}^N$ & fMRI signal at time $t$\\
$\overline{x}_t$ & Average of $x_{t,i}$ over ROIs\\
$\sigma_{t}$ & Standard deviation of $x_{t,i}$ over ROIs\\
$K$ & Number of discrete states \\ 
$\bm{c}_{\ell}  \in \mathbb{R}^N$ & Centroid of the $\ell$th cluster, where $\ell \in \{1,2, \ldots, K\}$\\
$L_t$ & Cluster label for $\bm{x}_t; L_t \in \{1, 2, \ldots, K\}$ \\
\hline
\end{tabular}
\label{tab:notation}
\end{table}

\subsubsection{K-means clustering} \label{k-means}

The K-means clustering is a simple and popular clustering method to partition the data points into $K$ mutually exclusive clusters. Various EEG and MEG microstate analysis \cite{michel2018eeg, zanesco2020within, tait2022meg} and the studies on temporal variability of functional connectivity states of fMRI data \cite{abrol2016chronnectome, abrol2017replicability, khosla2019machine} used the K-means clustering. It starts with a predefined number of clusters, $K$. We initialize the centroids of the clusters by the k-means++ algorithm~\cite{arthur2007k}. The k-means++ algorithm consists of the following steps. In step (i), we select one centroid uniformly at random from all the data points. In step (ii), for each data point $\mathbf{x}_t$ that is not yet selected as a centroid, we calculate the distance to the nearest centroid. In step (iii), we sample one $\mathbf{x}_t$ that is not yet selected as a centroid yet with the probability proportional to the square of the distance between $\mathbf{x}_t$ to the nearest centroid. In step (iv), we add the $\mathbf{x}_t$ sampled in the third step as a new centroid. We repeat steps (ii), (iii), and (iv) until we obtain $K$ centroids. This initialization method accelerates the convergence of the algorithm. 

Then, we refine the $K$ centroids, denoted by $\mathbf{c}_1$, $\ldots$, $\mathbf{c}_K$, as follows. The first step is to assign each data point $\mathbf{x}_t$ to the nearest centroid, i.e., the centroid realizing
\begin{equation} \label{eqn:mean_exp}
L_t= \text{arg} \min_{\ell \in \{1, \ldots, K\}} \left\| \mathbf{x}_t- \mathbf{c}_\ell \right\|,
\end{equation}
where $\left\| \cdot \right\|$ denotes the Euclidean norm.
The second step is to update the centroid of each cluster $\ell$ by the average of the data points belonging to the cluster as follows:
\begin{equation} \label{eqn:mean_max}
\mathbf{c}_\ell = \frac{\sum_{t=1}^T \mathbf{x}_t \delta_{L_t, \ell}}{\sum_{t=1}^T  \delta_{L_t, \ell}},
\end{equation}
where $\delta_{L_t, \ell}$ is the Kronecker delta; $\delta_{L_t, \ell}=1$ if $L_t = \ell$, and $\delta_{L_t, \ell} = 0$ otherwise. The Kronecker delta in the equation allows us to take the summation only over the data points belonging to the $\ell$th cluster. We repeat the first and second steps until the change in the residual sum of squares (RSS), defined by
\begin{equation} \label{eqn:k-means_conv}
\text{RSS}  = \sum_{t=1}^T \sum_{\ell=1}^K \delta_{L_t, \ell} \left\| \mathbf{x}_t- \mathbf{c}_\ell \right\|^2,
\end{equation}
between the two subsequent steps falls below $10^{-5}$ for the first time. We use the implementation of $k$-means in scikit-learn \cite{pedregosa2011scikit}.

\subsubsection{K-medoids clustering} \label{k-med}

The K-medoids clustering algorithm~\cite{kaufman2009finding} is a variant of the K-means clustering. The K-medoids clustering uses the original data points as the centroids of the clusters, referred to as medoids. In contrast, the K-means clustering uses the average of the points in the cluster as the centroid of the cluster. The K-medoids clustering begins with a set of $K$ data points as medoids, which we select using the k-medoids++ method. In fact, k-medoids++ is the same as k-means++. In the next step, we assign each $\mathbf{x}_t$ to the $\ell$th cluster whose medoid is closest to $\mathbf{x}_t$ in terms of the Euclidean distance. Then, we update the medoid of each cluster to $\mathbf{x}_t$ that belongs to the cluster and minimizes the sum of the Euclidean distance to the other data points in the same cluster. We repeat the last two steps until the dissimilarity score (i.e., the sum of the Euclidean distance from the medoid to the other data points in the cluster) stops changing for each cluster. We use the $k$-medoids implemented in scikit-learn~\cite{pedregosa2011scikit}.

\subsubsection{Agglomerative hierarchical clustering} \label{AC}

Agglomerative hierarchical clustering, which we simply call the agglomerative clustering (AC), is a bottom-up clustering method. The AC method initially regards each data point as a single-node cluster. Then, one merges a pair of clusters one after another based on a linkage criterion. Among various linkage criteria, we use the Ward's method implemented in scikit-learn~\cite{pedregosa2011scikit}. In each step of merging two clusters, the Ward's method minimizes the within-cluster variance, i.e., the squared Euclidean distance between $\mathbf{x}_t$ and the centroid of the new cluster to which $\mathbf{x}_t$ belongs, which is summed over $t \in \{1, \ldots, T \}$. We stop the cluster merging procedure once the number of clusters is equal to $K$.

\subsubsection{Atomize and agglomerate hierarchical clustering}

The Atomize and Agglomerate Hierarchical Clustering (AAHC) is another bottom-up hierarchical clustering algorithm \cite{tibshirani2005cluster, murray2008topographic, khanna2014reliability}. A main difference between AAHC and traditional bottom-up hierarchical clustering methods is that AAHC atomizes the worst cluster. In other words, AAHC disintegrates the worst cluster and assigns each member of this cluster to a different cluster instead of merging the entire worst cluster with the most similar cluster. 

AAHC uses the global explained variance (GEV) as a measure of the quality of the cluster \cite{murray2008topographic, khanna2014reliability, michel2018eeg,  poulsen2018microstate}. The GEV for the $\ell$th cluster is defined by 
\begin{equation} \label{eqn:gev}
\text{GEV}_\ell = \frac{ \sum_{t=1}^T \delta_{L_t, \ell} \, \text{corr}(\mathbf{x}_t, \mathbf{c}_\ell)^2 \, \sigma_t^2}{\sum_{t=1}^T\sigma_{t}^2},
\end{equation}
where $\text{corr}(\mathbf{x}_t, \mathbf{c}_\ell)$ is the cosine similarity between $\mathbf{x}_t $ and $\mathbf{c}_t$ given by 
\begin{equation} \label{eqn:correlation}
\text{corr}(\mathbf{x}_t, \mathbf{c}_\ell) = \frac{\langle \mathbf{x}_t, \mathbf{c}_\ell \rangle }{\left\| \mathbf{x}_t \right\| \left\| \mathbf{c}_\ell \right\|}.
\end{equation}
In Eq.~\eqref{eqn:correlation}, $\langle \mathbf{x}_t, \mathbf{c}_\ell \rangle$ is the inner product of $\mathbf{x}_t$ and $\mathbf{c}_\ell$. Variable $\sigma_t$ represents the standard deviation of the data point $\mathbf{x}_t$ across the ROIs and is given by Eq.~\eqref{eqn:gfp}. Quantity $\sigma_t$ is known as global field power (GFP) in the literature of microstate analysis for EEG and MEG data \cite{pascual1995segmentation, murray2008topographic, khanna2015microstates, tait2022meg}. For the second and third treatments of the global signal removal, it holds true that $\sigma_t = 1$ for any $t$ because of the global signal removal carried out in the last step of the treatment.

In the AAHC, we define the worst cluster as the one with the smallest $\text{GEV}_\ell$, $\ell \in\{1,2,\cdots, K\}$ and atomize it. Then, we assign each data point $\mathbf{x}_t$ of the atomized cluster to the $\ell$th cluster that maximizes Eq.~\eqref{eqn:correlation} \cite{murray2008topographic, khanna2014reliability}. As in the AC, the AAHC initially regards each $\mathbf{x}_t$ as a single-node cluster. We repeat finding the worst cluster, atomizing it, and assigning each $\mathbf{x}_t$ in the atomized cluster to a different cluster until the number of clusters reaches $K$.

\subsubsection{Topographic atomize and agglomerate hierarchical clustering}\label{TAAHC}

The Topographic Atomize and Agglomerate Hierarchical Clustering (TAAHC) is a modification of AAHC \cite{khanna2014reliability, poulsen2018microstate}. The difference between AAHC and TAAHC is that TAAHC defines the worst cluster to be the $\ell$th cluster that is the smallest in terms of the sum of the correlation of the data points in the cluster with its centroid $\mathbf{c}_\ell$ \cite{ khanna2014reliability, poulsen2018microstate}. In other words, the worst cluster $\ell$ is the minimizer of
\begin{eqnarray}\label{taahc:sumcorr}
\text{CRS}(\ell) = \sum_{t=1}^T \delta_{L_t, \ell}  \, \text{corr}(\mathbf{x}_t, \mathbf{c}_\ell) = \sum_{t=1}^T  \frac{\delta_{L_t, \ell} \, \langle \mathbf{x}_t, \mathbf{c}_{\ell} \rangle } {\left\|\mathbf{x}_t \right\|  \left\|\mathbf{c}_\ell \right\|}
\end{eqnarray}
over $\ell \in \{1, \ldots, K\}$. As in the AC and AAHC, the TAAHC first regards each $\mathbf{x}_t$ as a single-node cluster. Second, we identify the cluster with the smallest $\text{CRS}(\ell)$. Third, we atomize the selected cluster and reassign each of its member $\mathbf{x}_t$ to the cluster whose centroid is the closest to $\mathbf{x}_t$ in terms of $\text{corr}(\mathbf{x}_t, \mathbf{c}_{\ell})$. We iterate the second and third steps until we obtain $K$ clusters.  
  
\subsubsection{Bisecting K-means clustering} \label{bisect k-means}

The bisecting K-means method combines the K-means clustering method and divisive hierarchical clustering \cite{steinbach2000comparison}. Initially, we let all data points form a single cluster. Then, we apply the K-means clustering with $K=2$ to partition the data points into two clusters, by following the procedure described in section~\ref{k-means}. Then, we select the cluster that has the larger value of the dissimilarity defined for the $\ell$th  cluster by 
\begin{equation} \label{bisec_kmeans}
\rm{SSE}_\ell = \sum_{t=1}^T \delta_{L_t, \ell}  \left\| \mathbf{x}_t- \mathbf{c}_\ell \right\|^2.
\end{equation}

Then, we run the K-means clustering on the selected cluster to split it into two clusters. We repeat selecting the cluster with the largest $\rm{SSE}_{\ell}$
and bisecting it until we obtain $K$ clusters. We use the implementation of the bisecting K-means in scikit-learn~\cite{pedregosa2011scikit}. 

\subsubsection{Gaussian mixture model} \label{gmm}

The Gaussian mixture model (GMM) represents each cluster as a multivariate Gaussian distribution. We denote by $\mathcal{N}(\mathbf{\mu}_{\ell}, \mathbf{\Sigma}_{\ell})$, with $\ell \in \{1, 2, \ldots, K\}$, the multidimensional Gaussian distribution with mean vector $\mathbf{\mu}_{\ell}$ and covariance matrix $\mathbf{\Sigma}_{\ell}$~\cite{bishop2006pattern, ezaki2021modelling}. The GMM is given by
\begin{equation} \label{gmm:marginal}
p(\mathbf{x}_t) = \sum_{\ell = 1}^K \pi_\ell \, \mathcal{N}(\mathbf{x}_t|\mathbf{\mu}_\ell, \mathbf{\Sigma}_\ell),
\end{equation}
where $\pi_{\ell}$ is the mixing weight, i.e., the probability that a data point originates from the $\ell$th multivariate Gaussian distribution. Note that
$\sum_{\ell=1}^K \pi_{\ell} = 1$. The likelihood function for the set of all the data points is given by 
\begin{equation} \label{gmm:likelihood}
p(\mathbf{x}_1, \ldots, \mathbf{x}_T) = \prod_{t=1}^T \sum_{\ell=1}^K \pi_\ell \, \mathcal{N}(\mathbf{x}_t|\mathbf{\mu}_\ell, \mathbf{\Sigma}_\ell).
\end{equation}
We infer the parameter values by maximizing the log-likelihood function using an expectation-maximization (EM) algorithm~\cite{lindsay1995mixture, bishop2006pattern, ezaki2021modelling}. We regard $\mathbf{\mu}_\ell$ as the centroid of the $\ell$th cluster. Because the GMM is a soft clustering method, we assign each time point $t$ to the $\ell$th cluster that maximizes
$\hat{\pi}_\ell \mathcal{N} (\mathbf{x}_t | \hat{\mathbf{\mu}}_\ell, \hat{\mathbf{\Sigma}}_\ell )$, where
$\hat{\pi}_\ell$, $\hat{\mathbf{\mu}}_\ell$, and $\hat{\mathbf{\Sigma}}_\ell$ are the obtained maximum likelihood estimator.
We use the GaussianMixture class in scikit-learn, which uses K-means clustering for initializing the parameters \cite{pedregosa2011scikit}.

Among the seven methods that we employ to cluster the fMRI data, the GMM is the only parametric model. All the other methods are non-parametric clustering methods.

\subsection{Evaluation of the clustering methods}

The number of microstates estimated for EEG and MEG data depends on studies~\cite{pascual1995segmentation, khanna2014reliability, michel2018eeg, zanesco2020within}. Studies on temporal dynamics of functional connectivity in fMRI data are also diverse with respect to the number of clusters \cite{abrol2016chronnectome, abrol2017replicability, khosla2019machine}. Therefore, we examine the number of states, $K$, from 2 to 10 for each clustering algorithm. To compare the quality of the different clustering methods, we use the GEV given by Eq.~\eqref{eqn:gev}. The GEV captures the amount of the data variance explained by the microstates' centroids, also called the global map, cluster map, microstate map, and template map~\cite{murray2008topographic,  khanna2014reliability,von2018eeg, tait2022meg}.  We calculate the total GEV as the sum of the GEV over all the states, i.e.,
\begin{equation}
\text{GEV}_{\text{total}} = \sum_{\ell=1}^K \text{GEV}_{\ell}
\end{equation}
and average it over all the sessions and participants. A large value of the $\text{GEV}_{\text{total}}$ suggests that the obtained clustering is of high quality.

We also measure the quality of the clustering methods using the within-cluster sum of squares (WCSS)~\cite{jain1988algorithms}, also known as the distortion measure ~\cite{bishop2006pattern}. The WCSS is defined by 
\begin{equation} 
\label{eqn:wcss}
\text{WCSS} = \sum_{t=1}^T \sum_{\ell=1}^K \delta_{L_t, \ell}  \left\| \mathbf{x}_t - \mathbf{c}_\ell \right\|^2.
\end{equation}
A small WCSS value indicates that the data points are tightly clustered and therefore the clustering is of high quality.

\subsection{Comparison of state-transition dynamics between different sessions}

\subsubsection{Observables for the state-transition dynamics}\label{subsec:micro param}

To test reproducibility of the fMRI state-transition dynamics across participants and sessions, we measure the following five observables for each session. These observables are often used in the analysis of microstate dynamics for EEG and MEG data~\cite{baker2014fast, khanna2014reliability, zanesco2020within, tait2022meg} and activity patterns for fMRI data ~\cite{liu2013decomposition, abrol2016chronnectome, abrol2017replicability}.

First, we use the centroid of each of the $K$ states as an observable. The centroid $\mathbf{c}_\ell$ of the $\ell$th state represents the set of data points which are assigned to the $\ell$th state. We remind that the centroid is an $N$-dimensional vector.

Second, the coverage time of the $\ell$th state is the number of times $t \in \{1, \ldots, T\}$ in which the $\ell$th state appears. We normalize the coverage time of each state by dividing it by the total observation time, $T$.

Third, we measure the frequency of appearance of each state. If the $\ell$th state starts and then lasts for some time steps before transiting to a different state, then we say that this is a unique appearance of $\ell$. That is, we count the consecutive appearances as one unique appearance. The frequency of appearance of $\ell$ is defined as the number of unique appearance divided by $T$.

Fourth, the average lifespan of the $\ell$th state is the time spent in a unique appearance of $\ell$ that is averaged over all unique appearances of $\ell$. The average lifespan of $\ell$ is equal to the coverage time divided by the number of unique appearance of $\ell$.

Fifth, we investigate the frequency of transitions from one state to another as follows. Let $n_{\ell \ell'}$ be the number of times with which the transition from the $\ell$th state to the $\ell'$th state occurs in the given session, where $\ell' \neq \ell$. We define the transition probability from $\ell$ to $\ell'$ by $p_{\ell \ell'} = n_{\ell \ell'} / \sum_{\ell^{\prime\prime}=1; \ell^{\prime\prime} \neq \ell}^K n_{\ell \ell^{\prime\prime}}$ and we set $p_{\ell \ell} = 0$. The $K \times K$ transition probability matrix is given by $P = (p_{\ell \ell'})$ with $\ell, \ell' \in \{1, \ldots, K\}$.

\subsubsection{Discrepancy measures for comparing the state-transition dynamics between two sessions}
\label{sub:discrepancy-between-sessions}
For examining the reproducibility of state-transition dynamics between sessions of the same participant and between different participants, we need to compare observables between pairs of sessions. To this end, we first need to find the best matching of the states between the two sessions. For $K\in \{2, \ldots, 8\}$, we assess all the possible pairwise matchings of the states between the two sessions. This entails exhaustively matching every permutation of the $K$ states of the one session with the states of the another session. The total number of such pairwise matchings is equal to $K!$. For each matching, we calculate the correlation between centroids $\mathbf{c}_\ell$ and $\mathbf{c}_{\ell'}$ of the matched states, i.e., $\ell$th state in the first session and the $\ell'$th state in the second session, by $\text{corr}(\mathbf{c}_\ell, \mathbf{c}_{\ell'})$, where $\text{corr}$ is defined in Eq.~\eqref{eqn:correlation}. We then average $\text{corr}(\mathbf{c}_\ell, \mathbf{c}_{\ell'})$ over all the $K$ matched pairs of states in the two sessions and call it the centroid similarity. We select the matching that maximizes the centroid similarity among the $K!$ matchings.

For $K=9$ and $K=10$, we cannot assess all possible pairwise matchings due to combinatorial explosion. Therefore, we use a greedy search to find an approximately optimal matching. First, we find the pair of the $\ell$th state in the first session and the $\ell'$th state in the second session that maximizes $\text{corr}(\mathbf{c}_\ell, \mathbf{c}_{\ell'})$. Second, we select one state from the remaining $K-1$ states in the first session and one state from the remaining $K-1$ states in the second session such that the correlation between the two centroids is the largest, and we pair them. We repeat this procedure until all the $K$ states are matched between the two sessions.

Once we have determined the final matching between the $K$ states in the first session and those in the second session, we use the centroid dissimilarity, defined as $1-(\text{centroid similarity})$, as a measure of discrepancy between the set of $K$ states in the two sessions. The centroid dissimilarity ranges between $0$ and $2$. It is equal to $0$ if and only if the set of the $L$ centroid positions is exactly parallel between the two sessions.

The centroid similarity, $\text{corr}(\mathbf{c}_\ell, \mathbf{c}_{\ell'})$, only compares the direction of the two centroids, $\mathbf{c}_\ell$ and $\mathbf{c}_{\ell'}$, from the origin.
Therefore, we also measured the discrepancy between the set of $K$ states in the two sessions based on the Euclidean distance between $\mathbf{c}_\ell$ and $\mathbf{c}_{\ell'}$, given by
\begin{equation} \label{eqn:dissim}
d(\mathbf{c}_{\ell}, \mathbf{c}_{\ell'}) = \left\| \mathbf{c}_\ell- \mathbf{c}_{\ell'} \right\|^2.
\end{equation}
In the verification analysis, we searched for the best matching of the $K$ states between the two sessions by
minimizing the average of $d(\mathbf{c}_{\ell}, \mathbf{c}_{\ell'})$ over the $K$ matched pairs of states instead of maximizing the average of $\text{corr}(\mathbf{c}_\ell, \mathbf{c}_{\ell'})$. 
Similar to the case of using $\text{corr}(\mathbf{c}_\ell, \mathbf{c}_{\ell'})$, we did so by the exhaustive search when $K \in \{ 2, \ldots, 8 \}$ and by the greedy algorithm when $K \in \{ 9, 10 \}$. The dissimilarity obtained using the average of $d$ is equal to $0$ if and only if the set of the $L$ centroid positions is the same between the two sessions, and its large value implies a large discrepancy between the two sessions in terms of the centroid position.

For the coverage time, frequency of appearance, and average lifespan of states, we compute the total variation (TV) to quantify the difference in the state-transition dynamics between two sessions. Let $Q_i(\ell)$ be the coverage time, frequency of appearance, or average lifespan for the $\ell$th state in session $i$. For the notational convenience, we assume without loss of generality that we have matched the $\ell$th state in session $i$ with the $\ell$th state in session $j$.  For the coverage time of the $\ell$th microstate, we use the normalized coverage time defined in section~\ref{subsec:micro param} as $Q_i(\ell)$. The TV is defined by
\begin{equation} \label{eqn:TV}
\delta(Q_i, Q_j) = \max_{\ell \in \{1,2, \ldots, K\}} \left| Q_i(\ell)-Q_j(\ell)\right|,
\end{equation}
where $Q_i= \{Q_i(1), \ldots, Q_i(K) \}$.

To quantify the difference between the transition probability matrices for two sessions $i$ and $j$, denoted by $P^{(i)} = \left( p^{(i)}_{\ell \ell'}\right)$ and $P^{(j)} = \left( p^{(j)}_{\ell \ell'} \right)$, respectively, where $\ell, \ell' \in \{1, \ldots, K\}$, we calculate the Frobenius distance given by
\begin{equation} \label{eqn:trandiff}
\left\| P^{(i)}-P^{(j)} \right\|_\text{F} = \sqrt{\sum_{\ell = 1}^K \sum_{\ell'=1}^K \left| P^{(i)}_{\ell \ell'}- P^{(j)}_{\ell \ell'} \right|^2}.
\end{equation}

\subsubsection{Permutation test} \label{subsub:perm-test}

We hypothesize that the state-transition dynamics estimated from fMRI data is more consistent between different sessions of the same participant than between different participants. To test this hypothesis, we compare the dissimilarity between two sessions originating from the same participant and the dissimilarity between two sessions originating from different participants. If the former is smaller than the latter, then the state-transition dynamics is more reproducible within a participant than between different participants, supporting the potential ability of state-transition dynamics to be used for individual fingerprinting.

We measure the dissimilarity between a given pair of sessions in terms of one of the five observables (i.e., centroid position, distribution of the coverage time, normalized frequency of appearance of states, distribution of the average lifespan, or the transition probability matrix). For each observable, we compare the within-participant dissimilarity and between-participant dissimilarity using the normalized distance ND combined with the permutation test \cite{bernardi2020geometry, liu2020geometric}, which we adapt here for our purpose. Denote by $q(p, s)$ one of the five observables for participant $p \in \{1, \ldots, N_{\rm p} \}$ and session $s \in \{1, \ldots, N_{\rm s} \}$, where $N_{\rm p} = 8$ is the number of participants, and $N_{\rm s} = 10$ is the number of sessions per participant.
We define the ND by
\begin{eqnarray} \label{eqn:nd}
\text{ND}(q) &=  \frac{ \frac{2}{N_\text{p}(N_\text{p}-1) N_\text{s}} 
\sum\limits_{\substack{s=1}}^{N_\text{s}}  
\sum\limits_{\substack{p=1}}^{N_\text{p}}   
\sum\limits_{\substack{p'=1}}^{p-1}  
\tilde{d}\left(q(p, s), q(p', s)\right)}
{\frac{2}{N_\text{p} N_\text{s}(N_\text{s}-1)} 
\sum\limits_{\substack{p=1}}^{N_\text{p}}  
\sum\limits_{\substack{s=1}}^{N_\text{s}}   
\sum\limits_{\substack{s'=1}}^{s-1}  
\tilde{d}\left(q(p, s), q(p, s')\right)} \nonumber\\
&= \frac{(N_\text{s}-1) 
\sum\limits_{\substack{s=1}}^{N_\text{s}}  
\sum\limits_{\substack{p=1}}^{N_\text{p}}   
\sum\limits_{\substack{p'=1}}^{p-1}  
\tilde{d}\left(q(p, s), q(p', s)\right)}
{(N_\text{p}-1) 
\sum\limits_{\substack{p=1}}^{N_\text{p}}  
\sum\limits_{\substack{s=1}}^{N_\text{s}}   
\sum\limits_{\substack{s'=1}}^{s-1}  
\tilde{d}\left(q(p, s), q(p, s')\right)},
\end{eqnarray}
where $\tilde{d}$ denotes the dissimilarity (i.e., the Euclidean distance, TV, or Frobenius norm, depending on the observable; see section~\ref{sub:discrepancy-between-sessions}) between two sessions. The prefactor on the right-hand side on the first line of Eq.~\eqref{eqn:nd} accounts for the normalization;
there are $\frac{N_\text{p} (N_\text{p}-1) N_\text{s}}{2}$ and $\frac{N_\text{p} N_\text{s} (N_\text{s}-1)}{2}$ terms in the summation on the numerator and denominator, respectively. Therefore, the numerator of the right-hand side on the first line of Eq.~\eqref{eqn:nd} represents the average dissimilarity between two sessions obtained from different participants. The denominator represents the average dissimilarity between two sessions obtained from the same participant. If the state-transition dynamics are more consistent among different sessions within the same participant than among different sessions of different participants, we expect that $\text{ND}(q) > 1$.

To statistically test the $\text{ND}(q)$ value, we ran a permutation test~\cite{maris2007nonparametric}. Specifically, we carried out the following steps.
\begin{itemize}
    \item [(i)] Shuffle the values of $q$ across all participants and sessions uniformly at random. This process is equivalent to applying a random permutation on $\{q(1,1), q(1,2), \ldots, q(N_{\text{p}}, N_{\text{s}}) \}$. We denote the $q$ value for the $s$th session for $p$th participant after the random permutation by $q'(p, s)$. Note that the $q'$ value originates from any of the $N_{\text{p}}$ participants with probability $1/N_{\text{p}}$ and any of the $N_{\text{s}}$ sessions with probability $1/N_{\text{s}}$.
    \item [(ii)] Calculate $\text{ND}(q')$.
\item[(iii)] Repeat steps (i) and (ii) $R$ times. We set $R=10^4$.
\item[(iv)] The permutation $p$-value is equal to the fraction of the runs among the $R$ runs in which the $\text{ND}(q')$ value is larger than the empirical $\text{ND}(q)$ value.
\end{itemize}

\section{Results}\label{sec:results}

\subsection{Choice of the global signal reduction and clustering methods}

\begin{figure}
    \centering
    \includegraphics[width=1.05\linewidth]{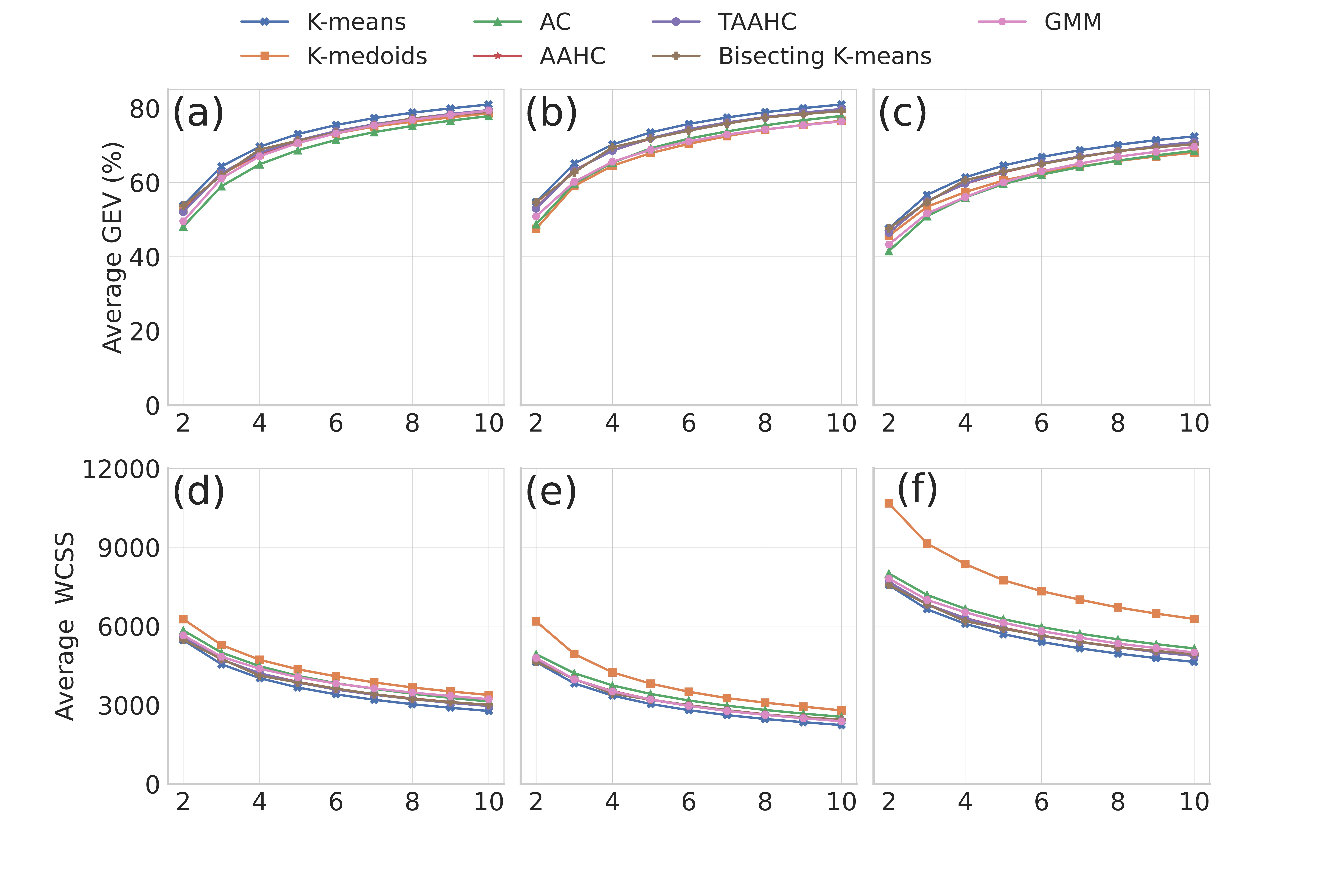} 
\caption{Performance of estimating discrete states from the DMN extracted from the MSC data. We show the results for the three treatments of global signal removal, seven clustering methods, and $K \in \{2, \ldots, 10 \}$.
(a)--(c): Total GEV. (d)--(f): WCSS. (a) and (d): First treatment of the global signal removal. (b) and (e): Second treatment. (c) and (f): Third treatment. Each $\text{GEV}_{\text{total}}$ and WCSS value shown is the average over the eight participants and ten sessions per participant. 
}
   \label{fig:gev_compare}
\end{figure}

We ran the seven clustering methods for each number of clusters, $K \in \{2, \ldots, 10 \}$, each of the ten sessions, each of the eight participants, and each of the three global signal removal methods for the DMN extracted from the MSC data. Then, we calculated the total GEV, i.e., $\text{GEV}_{\text{total}}$, for each combination of these variables as a measure of the quality of clustering. We show the $\text{GEV}_{\text{total}}$ values averaged over all the participants and sessions in Figure~\ref{fig:gev_compare}(a)--(c), for each combination of these variations. Each panel of Figure~\ref{fig:gev_compare} corresponds to a treatment of the global signal removal. In all the cases, $\text{GEV}_{\text{total}}$ increases as $K$ increases. We also find that $\text{GEV}_{\text{total}}$ is notably larger with the first and second treatments than the third treatment for all the seven clustering methods and that $\text{GEV}_{\text{total}}$ is slightly larger under the second than the first treatment for all values of $K$ and for all the clustering methods. 
Because the second treatment of the global signal removal shows the best performance in terms of clustering quality (i.e., providing the largest $\text{GEV}_{\text{total}}$), 
we use the second treatment in the following analyses. 

\begin{figure}
    \centering
    \includegraphics[width=1.05\linewidth]{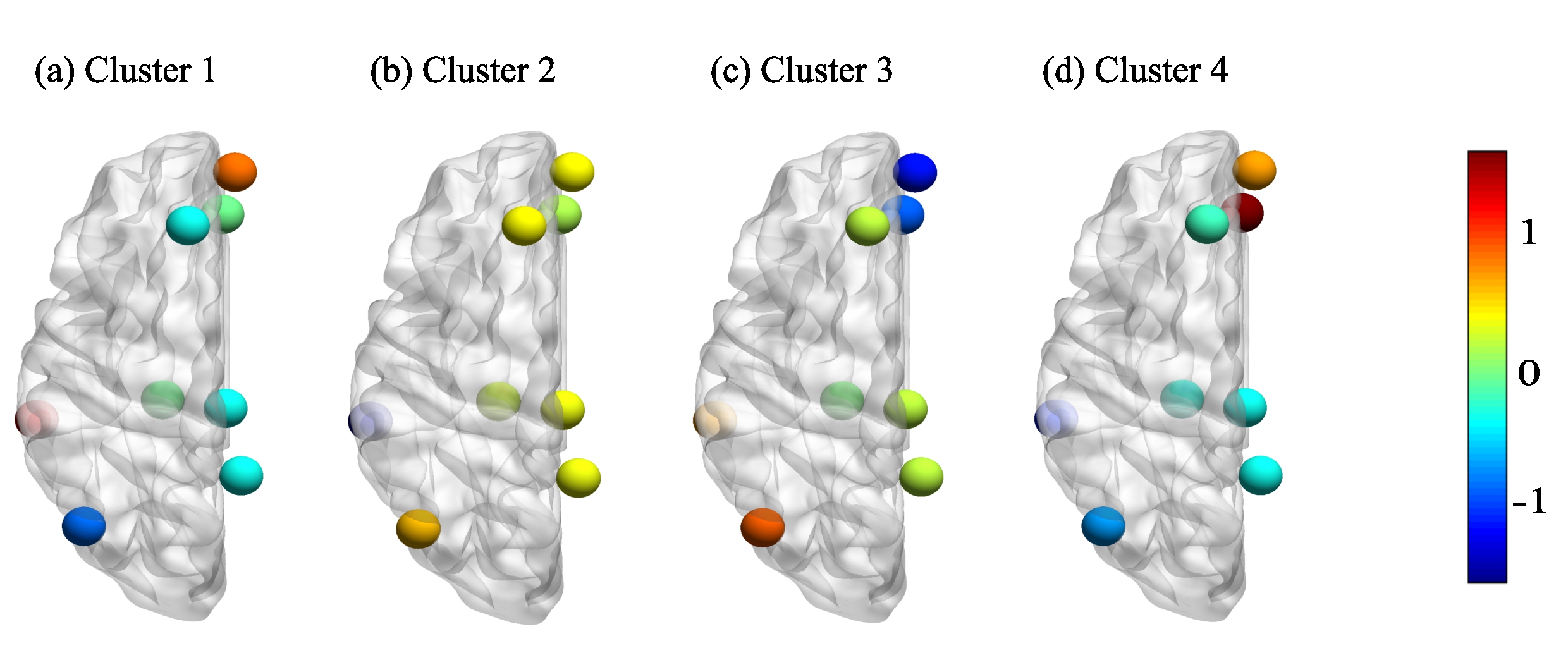} 
\caption{Brain map for each of the $K=4$ discrete states estimated by K-means for the first session of the first participant (i.e., MSC01). This choice of session and participant is arbitrary. We employed the second treatment of the global signal removal. Each circle represents an ROI in the left hemisphere. The color represents the activity averaged over the volumes belonging to the corresponding state. We used the BrainNet Viewer tool~\cite{xia2013brainnet} for the visualization.}
   \label{fig:brain-map}
\end{figure}

We select the three clustering methods with the largest $\text{GEV}_{\text{total}}$, which are the K-means, TAAHC, and bisecting K-means. For these three clustering methods, $\text{GEV}_{\text{total}}$ is around 70\% with $K=4$  (K-means: $70.22\pm 2.76\%$ (average $\pm$ standard deviation calculated on the basis of the 80 sessions of the MSC data), TAAHC: $68.56\pm 2.98\%$, bisecting K-means: $69.48\pm2.93\%$) and more than 75\% with $K=7$ (K-means: $77.49\pm 2.07\%$, TAAHC: $76.16\pm 2.29\%$, bisecting K-means: $75.85\pm2.31\%$). We show a brain map of each of the $K=4$ discrete states estimated by K-means in Figure~\ref{fig:brain-map}. As references, previous microstate analyses on EEG data found the $\text{GEV}_\text{total}$ values of $70.92\pm 3.65\%$ \cite{khanna2014reliability} and $65.80 \pm 4.90\%$~\cite{von2018eeg} using the K-means clustering, and $69.93\pm3.58\%$ using TAAHC
 \cite{khanna2014reliability}, all with $K=4$. Furthermore, $\text{GEV}_\text{total}$ values of $65.03\pm 6.13\%$ and $60.99\pm 5.62\%$ with $K=5$ were reported for EEG data recorded under eyes-closed and eyes-open conditions, respectively~\cite{zanesco2020within}. A MEG study reported a $\text{GEV}_\text{total}$ of $63.97 \pm 0.64\%$ using the K-means clustering with $K=10$~\cite{tait2022meg}. Our present data analysis with the fMRI data has yielded somewhat larger $\text{GEV}_\text{total}$  values than these studies. 

The GEV is based on the similarity in the direction of the $N$-dimensional fMRI signal, $\mathbf{x}_t$, and the centroid of the cluster, $\mathbf{c}_{L_t}$, where we remind that $L_t$ is the index of the cluster to which $\mathbf{x}_t$ belongs. Therefore, the GEV can be large even if $\mathbf{x}_t$ and $\mathbf{c}_{L_t}$ are not close to each other. Therefore, we also computed the WCSS, which is the sum of the distance between $\mathbf{x}_t$ and $\mathbf{c}_{L_t}$ over all the volumes.
We confirmed that the dependence of the WCSS on the global signal removal method, clustering method, and $K$
is similar to that with $\text{GEV}_{\text{total}}$ (see Figure~\ref{fig:gev_compare}(d)--(f)). Note that a large $\text{GEV}_{\text{total}}$ value implies a good clustering result, whereas a small WCSS value implies a good clustering result. In particular, with the WCSS, the second treatment of the global signal removal is the best among the three treatments, and the best three clustering methods remain the same, while the GMM performs as equally well as the TAAHC and the bisecting K-means for the second treatment of the global signal removal (see Figure~\ref{fig:gev_compare}(e)).  Therefore, in the remainder of this paper, we further focus our analysis only on the K-means, TAAHC, and bisecting K-means clustering methods.

Global signals can show spatio-temporal patterns and may provide specific pathological or psychological information~\cite{zhang2022beyond, wang2023frequency}. Therefore, we examined the quality of clustering produced by seven clustering methods and for $K\in\{2, \cdots, 10\}$ in the absence of global signal removal. The quality of clustering was worse without global signal removal than with global signal removal in terms of both $\text{GEV}_{\text{total}}$ and WCSS (see section S2). Therefore, we do not consider the analysis without global signal removal in the following sections.

\subsection{Test-retest reliability of the observables of the state-transition dynamics} \label{sub:result-observables}

We calculated the five observables, i.e., centroid of the clusters, coverage time, frequency, average lifespan, and transition probability matrix, of the estimated state-transition dynamics for each of the three selected clustering methods, each $K$ value, session, and participant. Then, we calculated the discrepancy in each observable between two sessions. To compare the state-transition dynamics of different sessions within the same participant, which we call the within-participant comparison, we calculated the discrepancy in terms of each observable for each pair of sessions for each participant. Because there are ten sessions for each of the eight participants, there are $\binom{10}{2} \times 8 = 360$ within-participant comparisons, where $\binom{}{}$ represents the binomial coefficient. To compare the state-transition dynamics between different participants, which we call the between-participant comparison, we calculated the discrepancy in terms of each observable between each pair of sessions obtained from different participants. There are $\binom{8}{2} \times 10 = 280$ between-participant comparisons. 

\begin{figure*}
\centering
\includegraphics[width=1\textwidth]{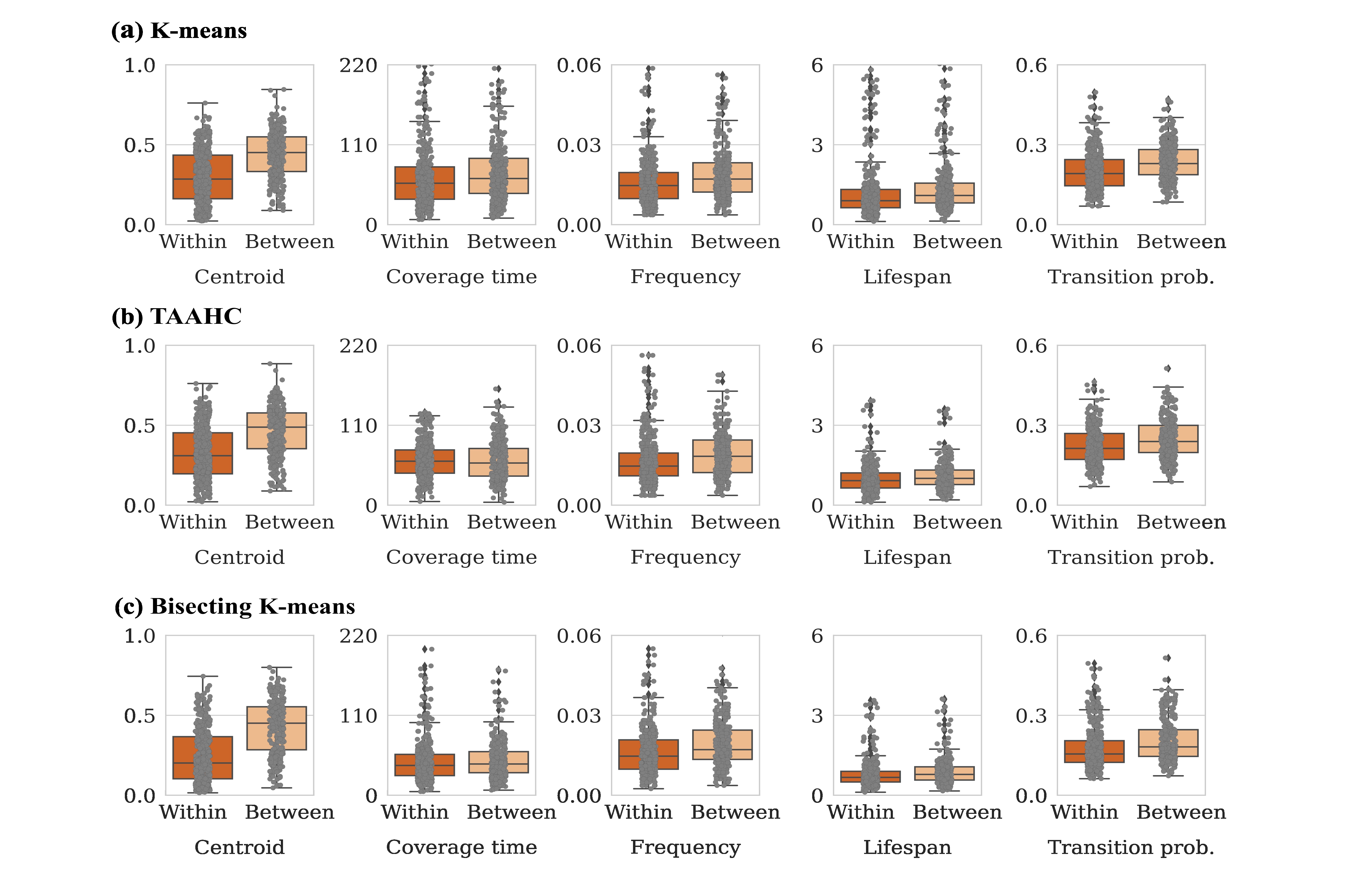}
\caption{Within-participant and between-participant reproducibility of the state-transition dynamics with $K=4$ states. (a) K-means. (b) TAAHC. (c) Bisecting K-means. ``Within'' and ``Between'' indicate the within-participant and between-participant comparisons, respectively. Each box plot shows the minimum, maximum, median, first quartile, and third quartile of the measurements. Each dot represents a session. ``Centroid'' abbreviates the centroid position, and ``Transition prob.'' abbreviates the transition probability matrix.}
\label{fig:compare-c4}
\end{figure*}

We show the distribution of the discrepancy measure for each observable with $K=4$, separately for the within-participant and between-participant comparisons, in Figure~\ref{fig:compare-c4}. Figures~\ref{fig:compare-c4}(a), \ref{fig:compare-c4}(b), and \ref{fig:compare-c4}(c) show the results for the K-means, TAAHC, and bisecting K-means, respectively. We find that the state-transition dynamics are visually more similar in the within-participant than between-participant comparisons across all the indices and for all the three clustering methods when we compare the minimum, maximum, median, first quartile, and third quartile values of each distribution. For all the three clustering methods, the gap between the within-participant and between-participant comparison is apparently the largest for the centroid position among the five observables. The gap between the within-participant and between-participant comparisons often looks subtle, in particular for the coverage time. The results with $K=7$ and $K=10$ are qualitatively the same as those with $K=4$ (see section S2).

To test the significance of the difference between the within-participant and between-participant session-to-session reproducibility of the state-transition dynamics, we carried out the permutation test. We computed the ND value for each clustering method, value of $K\in \{2, \ldots, 10\}$, and observable. Furthermore, we computed the ND values for $10^4$ randomized session-to-session comparisons. The permutation test concerns whether the ND value for the original session-to-session comparisons is significantly different from the ND values for the comparisons between the randomized pairs of sessions.

\begin{figure*}
\centering
\includegraphics[width=1\textwidth]{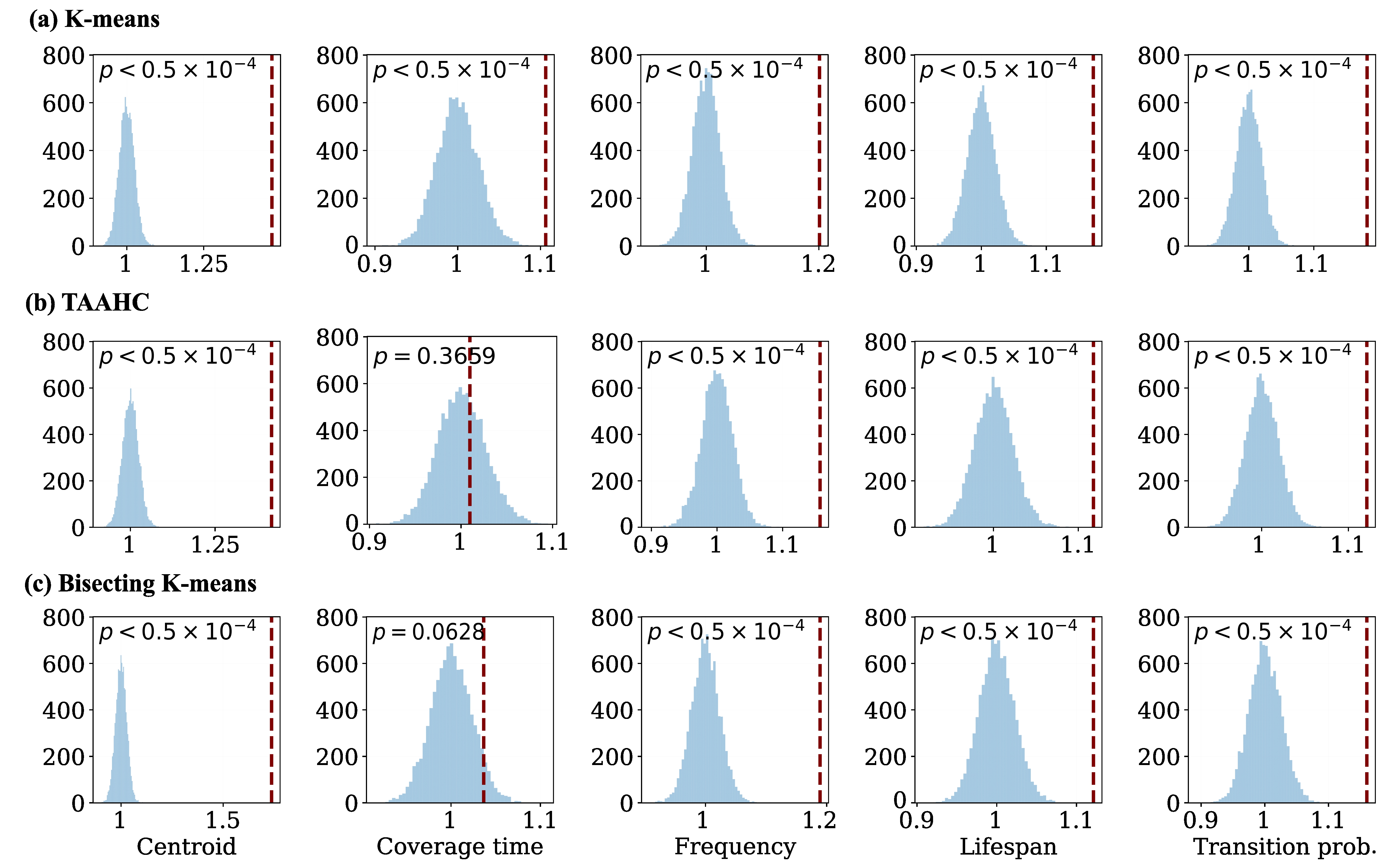}
\caption{Distribution of the ND values for the original and randomized session-to-session comparisons with $K=4$. (a) K-means. (b) TAAHC. (c) Bisecting K-means. The $p$ values shown are the uncorrected values.}
\label{fig:nd_all_c4}
\end{figure*}

With $K=4$, we show the ND value for the original session-to-session comparisons and the distribution of the ND values for the randomized sessions 
in Figure~\ref{fig:nd_all_c4}. Each panel of Figure~\ref{fig:nd_all_c4} shows a combination of the clustering method and the observable.
The vertical dashed lines represent the ND values for the original session-to-session comparisons.
We find that the result of the permutation test is significant in many cases even after correcting for multiple comparisons over the three clustering methods, nine values of $K$, and five observables; see the uncorrected $p$ values in the figure; an uncorrected $p = 0.00037$ corresponds to a Bonferroni corrected $p=0.05$. A small $p$ value implies that the within-participant session-to-session reproducibility is higher than the between-participant session-to-session reproducibility, suggesting the possibility of using the observable for fingerprinting individuals.

We tabulate the $p$ values from the permutation test for the three clustering methods, $K\in \{2, \ldots, 10 \}$, and the five observables in Table~\ref{p-values}. In the table, $p < 10^{-4}$ indicates that the ND value for the original session-to-session comparisons is farther from $1$ than all the $10^4$ randomized comparisons. The table shows that a majority of the $p$ values (i.e., 126 out of 135; 93.33\%) are smaller than 0.05 (shown with *). 
One hundred and seventeen of them (i.e., 86.67\% of the 135 comparisons) remain significant after the Bonferroni correction (shown with ***; equivalent to $p<0.00037$, uncorrected). Because there are 135 comparisons in the table and using the Bonferroni correction may be too stringent, we also counted the cases in which the uncorrected $p$ value is less than $0.001$, shown with **; there are 119 out of the 135 comparisons (i.e., 88.15\%) with $p<0.001$. We find that the number of significant $p$ values with the K-means and bisecting K-means is somewhat larger than with the TAAHC (with $p<0.05$, K-means, TAAHC, and bisecting K-means have 44, 39, and 43 significant comparisons, respectively). We also find that the $p$ values considerably depend on the observables. The permutation test result is strongly significant (i.e., $p<10^{-4}$) for all the clustering methods and $K$ values for the centroid position. In contrast, the number of significant combinations of the clustering method and $K$ value is smallest for the coverage. Lastly, we do not observe a notable dependence of the permutation test result on $K$.

\subsection{Robustness tests}

For validation, we also estimated the state-transition dynamics for the whole-brain network extracted from the MSC data. We show the results for the permutation test 
in Table~\ref{msc-whole-brain-p-values}. The results are similar to those for the DMN. In particular, the centroid position is the most effective among the five observables at distinguishing between the within-participant and between-participant comparisons, and the coverage is the least effective.

To examine the robustness of the proposed method with respect to fMRI experiments, we also ran the same permutation test for a whole-brain network obtained from the resting-state HCP data. Note that, among various parameters, the TR for the HCP data (i.e., 0.72 s) is substantially different from that for the MSC data (i.e., 2.2 s). It should also be noted that we use different atlases for the MSC and HCP data. While this decision is convenient for us because we have been using the obtained ROI-based fMRI data in different projects, it allows us to investigate the robustness of the proposed methods with respect to the atlas. The results, shown in Table~\ref{hcp-p-values}, are similar to those for the DMN and whole-brain networks extracted from the MSC data. However, the permutation test results were stronger for the HCP than MSC data (i.e., a larger number of significant comparisons among the 135 comparisons). In particular, the frequency, lifespan, and the transition probability matrix as well as the centroid position yielded the smallest possible $p$ value (i.e., $p <  10^{-4}$) for all pairs of the clustering method and $K$ value.

\begin{table*}
\centering
\caption{Results of the permutation test for the DMN extracted from the MSC data.  *: $p<0.05$, uncorrected; **: $p<0.001$, uncorrected; ***: $p<0.00037$, uncorrected (which is equivalent to $p<0.05$, Bonferroni corrected).  We remark that ``Centroid'' and "Trans. prob." abbreviate the centroid's position and the transition probability matrix, respectively.}
\begin{tabular}{>{\centering\arraybackslash}m{0.5cm} >{\centering\arraybackslash}m{0.5cm} >{\centering\arraybackslash}m{1.6cm} >{\centering\arraybackslash}m{1.6cm} >{\centering\arraybackslash}m{1.6cm} >{\centering\arraybackslash}m{1.6cm} >{\centering\arraybackslash}m{1.8cm}}
   \hline
   & $K$ & Centroid & Coverage & Frequency & Lifespan & Trans. prob. \\ 
   \hline                   
\multirow{9}{*}{\begin{turn}{90}K-means\end{turn}} 
				& 2  & $< 10^{-4***}$ & 0.1975 & $< 10^{-4***}$  &$< 10^{-4***}$& $< 10^{-4***}$   \\
                                  & 3  & $< 10^{-4***}$ & $< 10^{-4***}$ & $< 10^{-4***}$ & $< 10^{-4***}$  & $< 10^{-4***}$  \\
                                  & 4  & $< 10^{-4***}$ & $< 10^{-4***}$  & $< 10^{-4***}$ & $< 10^{-4***}$   & $< 10^{-4***}$    \\
                                  & 5  & $< 10^{-4***}$  & $< 10^{-4***}$ & $< 10^{-4***}$ & $< 10^{-4***}$  & $< 10^{-4***}$    \\
                                  & 6  & $< 10^{-4***}$  & $< 10^{-4***}$ & $< 10^{-4***}$ & $< 10^{-4***}$  & $< 10^{-4***}$    \\
                                  & 7  & $< 10^{-4***}$  & $< 10^{-4***}$ & $< 10^{-4***}$ & $< 10^{-4***}$  & $< 10^{-4***}$    \\
                                  & 8  & $< 10^{-4***}$  & $< 10^{-4***}$ & $< 10^{-4***}$ & $< 10^{-4***}$  & $< 10^{-4***}$    \\
                                  & 9  & $< 10^{-4***}$  & $< 10^{-4***}$ & $< 10^{-4***}$ & $< 10^{-4***}$  & $< 10^{-4***}$    \\
                                  & 10 & $< 10^{-4***}$  & $< 10^{-4***}$ & $< 10^{-4***}$ & $< 10^{-4***}$  & $< 10^{-4***}$    \\
                                  \hline 
\multirow{9}{*}{\begin{turn}{90}TAAHC\end{turn}}  
				& 2  & $< 10^{-4***}$ & $0.1077$ & $0.0001^{***}$  & $< 10^{-4***}$   & $0.0001^{***}$   \\
                                  & 3  & $< 10^{-4***}$ & $0.0001^{***}$  & $< 10^{-4***}$  & $< 10^{-4***}$   & $< 10^{-4***}$            \\
                                  & 4  & $< 10^{-4***}$ & $0.3659$      & $< 10^{-4***}$         & $< 10^{-4***}$ & $< 10^{-4***}$    \\
                                  & 5  & $< 10^{-4***}$    & $0.0928$         &  $< 10^{-4***}$  & $< 10^{-4***}$    & $< 10^{-4***}$     \\
                                  & 6  & $< 10^{-4***}$    & $0.0162^{*} $          & $< 10^{-4***}$         & $< 10^{-4***}$   & $< 10^{-4***}$   \\
                                  & 7  & $< 10^{-4***}$    &  0.5488           & $< 10^{-4***}$       & $0.0003^{***}$   & $< 10^{-4***}$  \\
                                  & 8  & $< 10^{-4***}$      &  $0.0642$      & $< 10^{-4***}$       & $0.0001^{***}$      & $< 10^{-4***}$  \\
                                  & 9  & $< 10^{-4***}$     & $0.0942$        &  $0.0001^{***}$        & $< 10^{-4***}$  & $< 10^{-4***}$  \\
                                  & 10 & $< 10^{-4***}$  & $0.0132^*$     & $< 10^{-4***}$ & $< 10^{-4***}$  &  $< 10^{-4***}$       \\
                                  \hline 
\multirow{9}{*}{\begin{turn}{90}Bisecting K-means\end{turn}} 
				& 2 & $< 10^{-4***}$  &  0.0568 &  $< 10^{-4***}$   & $< 10^{-4***}$& $< 10^{-4***}$   \\
                                  & 3  & $< 10^{-4***}$ & $< 10^{-4***}$ & $< 10^{-4***}$  & $< 10^{-4***}$ & $< 10^{-4***}$               \\
                                  & 4  & $< 10^{-4***}$  &  0.0628   &$< 10^{-4***}$   & $< 10^{-4***}$     & $< 10^{-4***}$       \\
                                  & 5  & $< 10^{-4***}$  & $0.0114^{*}$ & $< 10^{-4***}$ & $< 10^{-4***}$   & $< 10^{-4***}$    \\
                                  & 6  & $< 10^{-4***}$  & $0.0102^{*}$   & $< 10^{-4***}$ & $0.0003^{***}$  & $< 10^{-4***}$      \\
                                  & 7  & $< 10^{-4***}$  & $0.0003^{***}$         & $< 10^{-4***}$        & $< 10^{-4***}$     & $< 10^{-4***}$ \\
                                  & 8  & $< 10^{-4***}$   & $0.0017^{*}$   & $0.0004^{**}$     & $< 10^{-4***}$      & $< 10^{-4***}$    \\
                                  & 9  & $< 10^{-4***}$   & $0.0112^{*}$  &  $0.0009^{**}$    & $< 10^{-4***}$   & $< 10^{-4***}$  \\
                                  & 10 & $< 10^{-4***}$  & $0.0162^{*}$          & $< 10^{-4***}$    & $< 10^{-4***}$    & $< 10^{-4***}$   \\                                  \hline     
\end{tabular}
\label{p-values}
\end{table*}

\begin{table*}
\centering
\caption{Results of the permutation test for the whole-brain network extracted from the MSC data.  *: $p<0.05$, uncorrected; **: $p<0.001$, uncorrected; ***: $p<0.00037$, uncorrected (which is equivalent to $p<0.05$, Bonferroni corrected).  We remark that ``Centroid'' and "Trans. prob." abbreviate the centroid's position and the transition probability matrix, respectively.}
\begin{tabular}{>{\centering\arraybackslash}m{0.5cm} >{\centering\arraybackslash}m{0.5cm} >{\centering\arraybackslash}m{1.6cm} >{\centering\arraybackslash}m{1.6cm} >{\centering\arraybackslash}m{1.6cm} >{\centering\arraybackslash}m{1.6cm} >{\centering\arraybackslash}m{1.8cm}}
   \hline
  & $K$ & Centroid & Coverage & Frequency & Lifespan & Trans. prob. \\ 
   \hline 
\multirow{9}{*}{\begin{turn}{90}K-means\end{turn}}  
	   & 2 & $< 10^{-4***}$ & $0.045^*$ & $< 10^{-4***}$ & $< 10^{-4***}$ & $< 10^{-4***}$ \\ 
        ~ & 3 & $< 10^{-4***}$ &$ 0.0193^*$ & $< 10^{-4***}$ & $< 10^{-4***}$ & $< 10^{-4***}$ \\ 
        ~ & 4 & $< 10^{-4***}$ & $0.0414^*$ & $< 10^{-4***}$ & $< 10^{-4***}$ & $< 10^{-4***}$ \\ 
        ~ & 5 & $< 10^{-4***}$  & $ < 10^{-4***} $& $< 10^{-4***}$ & $< 10^{-4***}$ & $< 10^{-4***}$ \\ 
        ~ & 6 & $< 10^{-4***}$  & $0.0002^{***} $& $< 10^{-4***}$ & $< 10^{-4***}$ & $< 10^{-4***}$ \\ 
        ~ & 7 & $< 10^{-4***}$ & $0.0008^{**}$ & $< 10^{-4***}$ & $< 10^{-4***}$ & $< 10^{-4***}$ \\ 
        ~ & 8 &  $< 10^{-4***}$  & $10^{-4***} $& $< 10^{-4***}$ & $< 10^{-4***}$ & $< 10^{-4***}$ \\ 
        ~ & 9 &  $< 10^{-4***}$  & $< 10^{-4***} $& $< 10^{-4***}$ & $< 10^{-4***}$ & $< 10^{-4***}$ \\ 
        ~ & 10 &  $< 10^{-4***}$  & $< 10^{-4***} $& $< 10^{-4***}$ & $< 10^{-4***}$ & $< 10^{-4***}$ \\  \hline 
\multirow{9}{*}{\begin{turn}{90}TAAHC\end{turn}}  
	  & 2 & $< 10^{-4***}$  & $0.1747$ & $< 10^{-4***}$ & $< 10^{-4***}$ & $< 10^{-4***}$ \\ 
        ~ & 3 & $< 10^{-4***}$ & $0.0167^*$ & $< 10^{-4***}$ & $< 10^{-4***}$ & $< 10^{-4***}$ \\ 
        ~ & 4 & $< 10^{-4***}$ & $0.4553$ & $< 10^{-4***}$ & $0.0001^{***}$ & $< 10^{-4***}$ \\ 
        ~ & 5 & $< 10^{-4***}$  & $0.0288^*$ & $0.0001^{***}$ & $< 10^{-4***}$ & $< 10^{-4***}$\\ 
        ~ & 6 & $< 10^{-4***}$ & $0.5063$ & $0.0002^{***}$ & $< 10^{-4***}$ & $< 10^{-4***}$ \\ 
        ~ & 7 & $< 10^{-4***}$ & $0.7551$ & $0.0008^{**}$ & $< 10^{-4***}$ & $< 10^{-4***}$ \\ 
        ~ & 8 & $< 10^{-4***}$ & 0.1887 & $0.0001^{***}$ & $< 10^{-4***}$ & $< 10^{-4***}$ \\ 
        ~ & 9 & $< 10^{-4***}$  & $0.0874$& $< 10^{-4***}$ & $< 10^{-4***}$ & $< 10^{-4***}$ \\ 
        ~ & 10 & $< 10^{-4***}$ & $0.0426^*$ & $< 10^{-4***}$ & $< 10^{-4***}$ & $< 10^{-4***}$ \\  \hline 
\multirow{9}{*}{\begin{turn}{90}Bisecting K-means\end{turn}} 
	   & 2 & $< 10^{-4***}$ &$0.0283^*$ & $< 10^{-4***}$ & $< 10^{-4***}$ & $< 10^{-4***}$ \\ 
        ~ & 3 & $< 10^{-4***}$  & $0.0597$ & $< 10^{-4***}$ & $0.0007^{**}$ & $0.0012^{*}$ \\ 
        ~ & 4 & $< 10^{-4***}$  & $0.0402^*$ & $< 10^{-4***}$ & $< 10^{-4***}$ & $< 10^{-4***}$ \\ 
        ~ & 5 & $< 10^{-4***}$ & 0.2698 & $0.0001^{***}$ & $0.0003^{***}$ & $0.065$ \\ 
        ~ & 6 & $< 10^{-4***}$ & $0.0341^*$ & $< 10^{-4***}$ & $< 10^{-4***}$& $< 10^{-4***}$ \\ 
        ~ & 7 & $< 10^{-4***}$  & $0.5185$ & $0.0002^{***}$ & $< 10^{-4***}$ & $< 10^{-4***}$ \\ 
        ~ & 8 & $< 10^{-4***}$  & $< 10^{-4***}$ & $< 10^{-4***}$ & $< 10^{-4***}$ & $< 10^{-4***}$ \\ 
        ~ & 9 & $< 10^{-4***}$  & 0.0886 & $0.0027^{*}$  & $< 10^{-4***}$& $< 10^{-4***}$ \\ 
        ~ & 10 & $< 10^{-4***}$ & $0.0012^{*}$ & $< 10^{-4***}$ & $< 10^{-4***}$ & $< 10^{-4***}$ \\ 
         \hline     
\end{tabular}
\label{msc-whole-brain-p-values}
\end{table*}

\begin{table*}
\centering
\caption{Results of the permutation test for the whole-brain network extracted from the HCP data.  *: $p<0.05$, uncorrected; **: $p<0.001$, uncorrected; ***: $p<0.00037$, uncorrected (which is equivalent to $p<0.05$, Bonferroni corrected).  We remark that ``Centroid'' and "Trans. prob." abbreviate the centroid's position and the transition probability matrix, respectively.}
\begin{tabular}{>{\centering\arraybackslash}m{0.5cm} >{\centering\arraybackslash}m{0.5cm} >{\centering\arraybackslash}m{1.6cm} >{\centering\arraybackslash}m{1.6cm} >{\centering\arraybackslash}m{1.6cm} >{\centering\arraybackslash}m{1.6cm} >{\centering\arraybackslash}m{1.8cm}}
   \hline
  & $K$ & Centroid & Coverage & Frequency & Lifespan & Trans. prob.\\ 
   \hline  
\multirow{9}{*}{\begin{turn}{90}K-means\end{turn}} 
	   & 2 & $< 10^{-4***}$ & $ 0.0308^{*}$ & $< 10^{-4***}$ & $< 10^{-4***}$ & $< 10^{-4***}$ \\
        ~ & 3 & $< 10^{-4***}$ &$ < 10^{-4***}$ & $< 10^{-4***}$ & $< 10^{-4***}$ & $< 10^{-4***}$ \\ 
        ~ & 4 & $< 10^{-4***}$  & $< 10^{-4***}$ & $< 10^{-4***}$ & $< 10^{-4***}$ & $< 10^{-4***}$ \\ 
        ~ & 5 & $< 10^{-4***}$ & $< 10^{-4***}$ & $< 10^{-4***}$ & $< 10^{-4***}$ & $< 10^{-4***}$ \\ 
        ~ & 6 & $< 10^{-4***}$  & $< 10^{-4***}$ & $< 10^{-4***}$ & $< 10^{-4***}$ & $< 10^{-4***}$ \\ 
        ~ & 7 & $< 10^{-4***}$  & $< 10^{-4***}$ & $< 10^{-4***}$ & $< 10^{-4***}$ & $< 10^{-4***}$ \\ 
        ~ & 8 & $< 10^{-4***}$  & $< 10^{-4***}$ & $< 10^{-4***}$ & $< 10^{-4***}$ & $< 10^{-4***}$ \\ 
        ~ & 9 & $< 10^{-4***}$ & $< 10^{-4***}$ & $< 10^{-4***}$ & $< 10^{-4***}$ & $< 10^{-4***}$ \\ 
        ~ & 10 & $< 10^{-4***}$ & $< 10^{-4***}$ & $< 10^{-4***}$ & $< 10^{-4***}$ & $< 10^{-4***}$ \\  \hline 
\multirow{9}{*}{\begin{turn}{90}TAAHC\end{turn}}  
	   & 2 & $< 10^{-4***}$ & $0.0001^{***}$  & $< 10^{-4***}$ & $< 10^{-4***}$ & $< 10^{-4***}$ \\ 
        ~ & 3 & $< 10^{-4***}$  & $0.0016^{*}$ & $< 10^{-4***}$ & $< 10^{-4***}$ & $< 10^{-4***}$ \\ 
        ~ & 4 & $< 10^{-4***}$  & $0.0305^{*}$ & $< 10^{-4***}$ & $< 10^{-4***}$ & $< 10^{-4***}$ \\ 
        ~ & 5 & $< 10^{-4***}$  & $0.0297^{*}$ & $< 10^{-4***}$ & $< 10^{-4***}$& $< 10^{-4***}$ \\ 
        ~ & 6 & $< 10^{-4***}$  & $0.0164^*$ & $< 10^{-4***}$ & $< 10^{-4***}$ & $< 10^{-4***}$ \\ 
        ~ & 7 & $< 10^{-4***}$  & $0.2864$ & $< 10^{-4***}$ & $< 10^{-4***}$ & $< 10^{-4***}$ \\ 
        ~ & 8 & $< 10^{-4***}$  & $0.0699$ & $< 10^{-4***}$ & $< 10^{-4***}$ & $< 10^{-4***}$ \\ 
        ~ & 9 & $< 10^{-4***}$ & $0.0203^{*}$ & $< 10^{-4***}$ & $< 10^{-4***}$ & $< 10^{-4***}$ \\ 
        ~ & 10 &  $< 10^{-4***}$ & $0.0764$ & $< 10^{-4***}$ & $< 10^{-4***}$ & $< 10^{-4***}$ \\ \hline 
\multirow{9}{*}{\begin{turn}{90}Bisecting K-means\end{turn}} & 2 & $< 10^{-4***}$  & $0.0015^{*}$ & $< 10^{-4***}$ & $< 10^{-4***}$ & $< 10^{-4***}$ \\ 
        ~ & 3 & $< 10^{-4***}$  & $0.0531$ & $< 10^{-4***}$ & $< 10^{-4***}$ & $< 10^{-4***}$ \\ 
        ~ & 4 & $< 10^{-4***}$ &  $0.0078^{*}$ & $< 10^{-4***}$ & $< 10^{-4***}$ & $< 10^{-4***}$ \\ 
        ~ & 5 & $< 10^{-4***}$  & $< 10^{-4***}$ & $< 10^{-4***}$ & $< 10^{-4***}$ & $< 10^{-4***}$ \\ 
        ~ & 6 & $< 10^{-4***}$  & $0.0009^{**}$ & $< 10^{-4***}$ & $< 10^{-4***}$ & $< 10^{-4***}$ \\ 
        ~ & 7 & $< 10^{-4***}$  & $0.005^{*}$ & $< 10^{-4***}$ & $< 10^{-4***}$ & $< 10^{-4***}$ \\ 
        ~ & 8 & $< 10^{-4***}$ & $< 10^{-4***}$ & $< 10^{-4***}$& $< 10^{-4***}$ & $< 10^{-4***}$ \\ 
        ~ & 9 &  $< 10^{-4***}$ & $< 10^{-4***}$ & $< 10^{-4***}$ & $< 10^{-4***}$ & $< 10^{-4***}$ \\ 
        ~ & 10 & $< 10^{-4***}$ & $< 10^{-4***}$ & $< 10^{-4***}$ & $< 10^{-4***}$ & $< 10^{-4***}$ \\
\hline     
\end{tabular}
\label{hcp-p-values}
\end{table*}

Moreover, as we noted earlier, our main definition of the centroid dissimilarity relies on the (dis)similarity between $\bm{c}_{L_t}$ and $\bm{x}_t$ only in terms of the direction. Therefore, we reran the permutation test by replacing the centroid (dis)similarity by the WCSS to measure the average distance between $\bm{c}_{L_t}$ and $\bm{x}_t$. This change not only affected the discrepancy measure between two sessions in terms of the centroid position but also the discrepancy between pairs of sessions in terms of the other four observables (i.e., coverage time, frequency of appearance of each state, average lifespan, and transition probability matrix). This is because changing the discrepancy measure for cluster centroids affects how the set of centroids (and therefore clusters) is matched between two given sessions. We confirmed that the permutation test results with the WCSS is similar to those with $\text{GEV}_{\text{total}}$ (see section S4). In particular, the $p$ values were overall small, the results tended to be more significant for the K-means and bisecting K-means than for the TAAHC, and for the centroid position and the transition probability matrix than for the other three observables.

Lastly, we employed a 59-ROI DMN~\cite{power2011functional} extracted from MSC data to further assess the impact of the dimension reduction. We found that results of the test-retest reproducibility are roughly as good as those with the 12-ROI DMN (see section S5).

\section{Discussion}\label{sec:discussion}

We carried out a comparative study of methods to cluster volumes of the fMRI to extract time series of the system's state, akin to microstate analysis for EEG and MEG data, for each recording session. We found that aggregating the symmetrically located ROIs into one ROI and then conducting the global signal removal yielded a high accuracy of clustering in terms of the total GEV and WCSS. We obtained total GEV values that are somewhat larger than those obtained in previous studies for EEG microstate analysis~\cite{khanna2014reliability, von2018eeg, zanesco2020within, tait2022meg}, which suggests that fMRI state-transition dynamics analysis may be promising. Furthermore, by carrying over the three clustering methods yielding the best clustering performance to a test-retest reliability analysis, we found that, for different fMRI data sets and different networks, test-retest reliability was higher in the within-participant comparison than the between-participant comparison. This result held true for most combinations of the number of clusters, $K \in \{2, \ldots, 10\}$, and index quantifying the estimated state-transition dynamics. We also found that the K-means clustering yielded the highest test-retest reliability among the three clustering methods. The present results suggest that clustering-based analysis of state-transition dynamics, which is substantially simpler than the hidden Markov model~\cite{baker2014fast, ryali2016temporal,  vidaurre2017brain, taghia2017bayesian, nielsen2018predictive, vidaurre2018spontaneous, vidaurre2021new,  ezaki2021modelling, tait2022tool, tait2022meg} and the energy landscape analysis~\cite{watanabe2013energy, watanabe2014energy, ezaki2017energy, ezaki2018ge}, may be a sufficiently competitive method to derive state-transition dynamics in fMRI data.

The microstate analysis was originally proposed for EEG data~\cite{lehmann1987eeg, koenig2002millisecond, khanna2015microstates, hatz2016reliability, michel2018eeg, von2018eeg, zhang2021reliability}. Microstates in EEG data are typically of the order of 100 ms. One cannot directly associate the discrete states estimated from fMRI data with EEG or MEG microstates because the time resolution of fMRI data is much lower than 100 ms; a typical TR is approximately between 1 to 3 seconds. Furthermore, the typical duration of a discrete state is longer than one TR. For example, the average lifespan of a state was $3.3$ TR, $2.5$ TR, and $2.2$ TR when we estimated four, seven, and ten states, respectively, for the DMN extracted from the MSC data. Therefore, cognitively or physiologically relevant discrete states estimated for fMRI data~\cite{allen2014tracking, abrol2017replicability, vidaurre2017brain} may be different  from those captured by microstates in EEG and MEG data. However, promising correspondences between EEG microstates and fMRI states have been reported \cite{britz2010bold, yuan2012spatiotemporal, allen2014tracking, abreu2021eeg}. Analyzing simultaneously recorded EEG-fMRI data may further reveal connection between EEG microstates and discrete states for fMRI data~\cite{britz2010bold, chang2013eeg, preti2014epileptic, ahmad2016simultaneous, keinanen2018fluctuations, brechet2019capturing, lurie2020questions, mulert2022simultaneous}.

We examined test-retest reliability of discrete states estimated by clustering activity pattern vectors of fMRI data. In contrast, various previous studies estimated discrete states by clustering functional networks from fMRI data  \cite{sakouglu2010method, hutchison2013resting, leonardi2013principal, calhoun2014chronnectome, baker2014fast, allen2014tracking, abrol2016chronnectome, ryali2016temporal, taghia2017bayesian, nielsen2018predictive}. Our methods of test-retest reliability analysis do not depend on how the discrete states are estimated and therefore are applicable to the case of state-transition dynamics of functional networks. To the best of our knowledge, no work has systematically compared the reliability between state-transition dynamics estimated for spatial activity patterns or their vectorized versions and those estimated for functional networks from the same fMRI data. Such a comparative analysis may better inform us whether activity patterns or functional networks are more powerful biomarkers than the other when combined with state-transition dynamics modeling. In a similar vein, the aforementioned studies pursuing similarity between EEG microstates and fMRI dynamic states have been confined to the case in which fMRI dynamic states are estimated from dynamics of functional connectivity, not dynamics of activity patterns. These topics warrant future work.

In EEG microstate analysis, it is common to generate global microstate maps, which is to determine a given number of microstates by clustering candidate EEG maps obtained from different participants altogether. Then, one matches the obtained microstate maps shared by all the participants to the individual EEG maps from the individual participants to determine the microstate dynamics for each participant~\cite{lehmann2005eeg, britz2010bold, khanna2014reliability, abrol2017replicability, zanesco2020within}. For EEG data, this approach has been shown to accrue higher reliability than microstate maps estimated separately for individual participants \cite{khanna2014reliability}. Nevertheless, we have estimated the states separately for individual participants (and for individual sessions) in the present study. This is because, for fMRI data, one often estimates state dynamics separately for each individual, which allows one to study subject variability of the estimated state dynamics or to exploit it~\cite{allen2014tracking, rashid2014dynamic, taghia2017bayesian, nielsen2018predictive}.
In contrast, pooling fMRI data from different participants to generate across-participant templates of discrete states is also a common practice~\cite{liu2013decomposition, rashid2014dynamic, allen2014tracking, abrol2017replicability, paakki2021co}. In fact, one can run our test-retest reliability analysis even if we estimate the templates of the discrete states shared by all participants, with the exception of the cluster centroid, $\bm{c}_{\ell}$, as an observable of the estimated state dynamics; if we use a shared template, $\bm{c}_{\ell}$ is the same for all sessions and individuals and therefore one cannot compare its reliability within versus between participants. We point out that comparison of the reliability between shared templates and individualized templates of discrete states for fMRI data, as was done for EEG data \cite{khanna2014reliability}, is underexplored.

We ran a permutation test to statistically compare the within-participant and between-participant test-retest reliability. This permutation test is an adaptation of what we recently developed for energy landscape analysis~\cite{khanra2023reliability} to the case of clustering-based state-transition dynamics. This method is not limited to fMRI data. It is straightforward to use it for EEG and MEG microstate data analysis obtained from multiple participants and multiple sessions per participant. Our code is publicly available on GitHub~\cite{SLIslam2023}. The only requirement is to define observables and to be able to measure the discrepancy in the observable between an arbitrary pair of sessions. Assessing test-retest reliability in EEG~\cite{khanna2014reliability, michel2018eeg, liu2020reliability, zhang2021reliability} and MEG~\cite{garces2016quantifying, candelaria2020reduced} data using this technique as well as furthering the application to fMRI data in health and disease may be fruitful.
 
\section*{Acknowledgements}

T.W. acknowledges support from the Japan Society for Promotion of Sciences (TW, 19H03535, 21H05679, 23H04217) 
N.M. acknowledges support from the Japan Science and Technology Agency (JST) Moonshot R\&D (under grant no.\,JPMJMS2021), the National Science Foundation (under grant no.\,2204936), and JSPS KAKENHI (under grant nos.\,JP 21H04595 and 23H03414).
Two publicly available data sets were used in this work. The first data set was provided by the Midnight Scan Club (MSC) project, funded by NIH Grants NS088590, TR000448 (NUFD), MH104592 (DJG), and HD087011 (to the Intellectual and Developmental Disabilities Research Center at Washington University); the Jacobs Foundation (NUFD); the Child Neurology Foundation (NUFD); the McDonnell Center for Systems Neuroscience (NUFD, BLS); the Mallinckrodt Institute of Radiology (NUFD); the Hope Center for Neurological Disorders (NUFD, BLS, SEP); and Dart Neuroscience LLC. This data was obtained from the OpenfMRI database. Its accession number is ds000224. The second data set was provided by the Human Connectome Project, WU-Minn Consortium (Principal Investigators: David Van Essen and Kamil Ugurbil; 1U54MH091657) funded by the 16 NIH Institutes and Centers that support the NIH Blueprint for Neuroscience Research; and by the McDonnell Center for Systems Neuroscience at Washington University.  The authors also acknowledge support provided by the Center for Computational Research at the University at Buffalo for the processing of HCP data.

\bibliographystyle{elsarticle-num}
\bibliography{fmri-bibliography.bib}

\newpage
\onecolumn
\begin{center}
{\Large Supplementary Information for \\
State-transition dynamics of resting-state functional magnetic resonance imaging data: Model comparison and test-to-retest analysis}\\
\vspace{12pt}
\end{center}

\begin{multicols}{2}
\renewcommand\thesection{S1}
\section{Correspondence between each AAL ROI and the brain system}\label{si:aal_roi}
As we described in section ``Human Connectome Project data", we mapped the 116 ROIs of the AAL atlas for the HCP data to different representative brain systems based on the Euclidian distance between the centroid of each ROI in the AAL atlas and the centroid of each ROI in the Schaefer atlas. Then, we excluded 42 ROIs labeled as ‘subcortical’ or ‘cerebellar’. With this exclusion, we consider the remaining seven brain systems: the control network, default mode network (DMN), dorsal attention network (DAN), limbic network, salience/ventral attention network, somatomotor network, and the visual network. We show the mapping between these brain systems and the AAL ROIs in Table~\ref{tab:aal_roi}.

\begin{table*}[htp]
\renewcommand\thetable{S1}
\centering
\caption{Mapping of AAL ROIs to the brain system}
\begin{tabular}{ l | l } 
 \hline
Brain system  & ROI in the AAL atlas\\ 
\hline
Control network & 8, 10, 13, 14, 32, 34, 61, 66, 88, 89 \\
DMN & 3,  4,  9, 15, 16, 23, 24, 25, 26, 31, 33, 65, 68, 82, 83, 84, 85, 86, 90 \\
DAN & 11, 12, 57, 59, 60, 62, 67, 69, 70 \\
Limbic network & 5,  6, 19, 20, 27, 28, 87 \\
Salience/ventral attention network & 1,  7, 29, 30, 63, 64\\
Somatomotor network & 2, 17, 18, 45, 46, 49, 50, 58, 79, 80, 81 \\
Visual network & 35, 36, 43, 44, 47, 48, 51, 52, 53, 54, 55, 56 \\
\hline
\end{tabular}
\label{tab:aal_roi}
\end{table*}

\renewcommand\thesection{S2}
\section{Quality of clustering without global signal removal}\label{si:without gs}

We evaluate the quality of clustering when the global signal removal is omitted as follows.
We consider the DMN extracted from the MSC, which has $\tilde{N} = 12$ ROIs,
and the symmetrized variant of the DMN, which has $\tilde{N} = 8$ ROIs and is obtained by averaging of the time series over each pair of symmetric ROIs from the left and right hemispheres.
The non-symmetrized DMN corresponds to the third treatment described in the main text. The symmetrized DMN corresponds to the first and second treatments. In this section, we use fMRI signals from these DMNs without global signal removal.

\begin{figure*}
\renewcommand\thefigure{S1}
\centering
\includegraphics[width=0.8\textwidth]{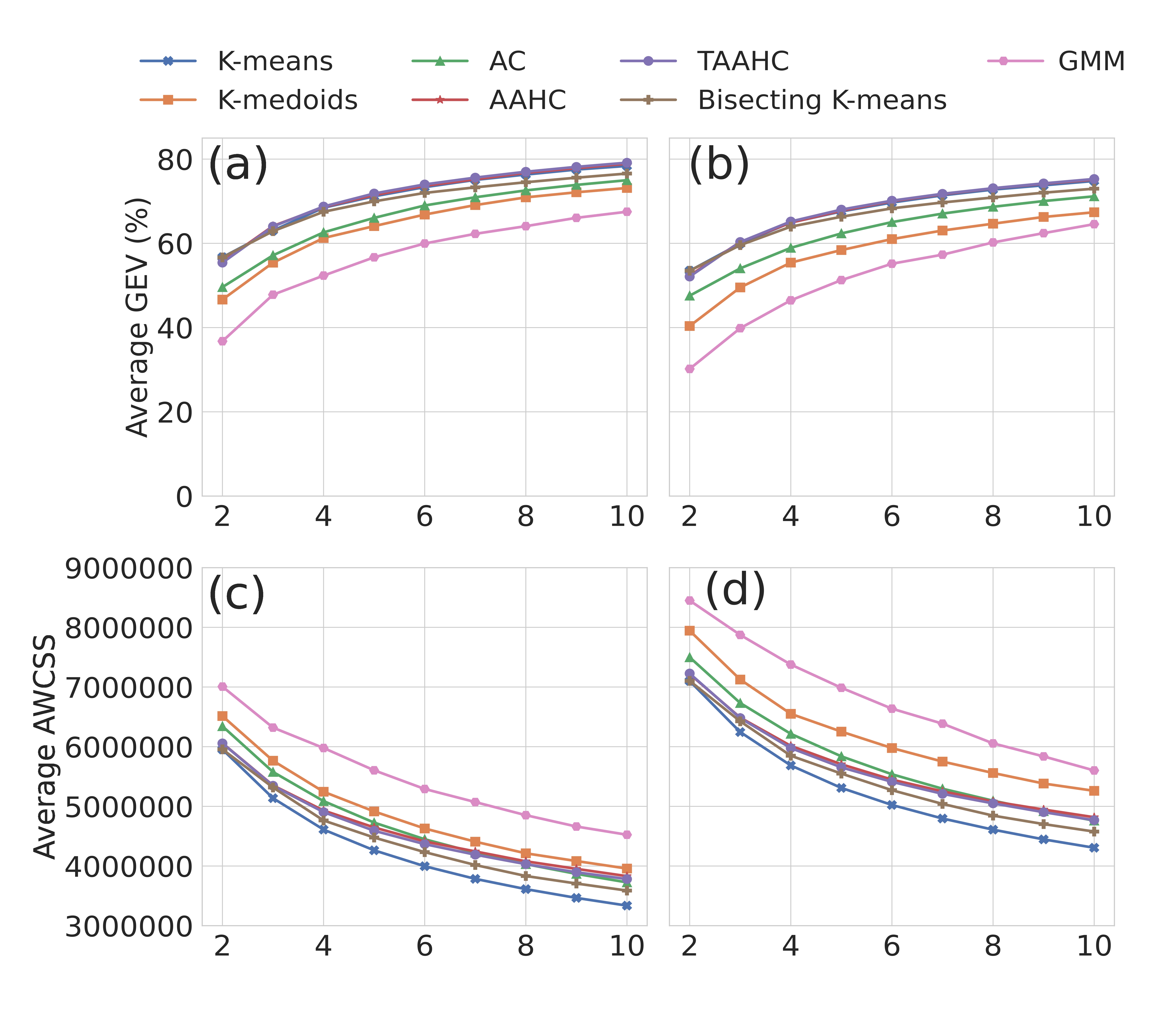}
 \caption{Performance of estimating discrete states without global signal removal. 
We use the DMN extracted from the MSC data. We show the results for the seven clustering methods and $K \in \{2, \ldots, 10 \}$.
(a) $\text{GEV}_{\text{total}}$, $\tilde{N}=8$ ROIs. (b) $\text{GEV}_{\text{total}}$, $\tilde{N}=12$. (c) WCSS, $\tilde{N}=8$. (d) WCSS, $\tilde{N}=12$. Each $\text{GEV}_{\text{total}}$ and WCSS value shown is the average over the eight participants and ten sessions per participant.
}
\label{fig:si_gev_compare}
\end{figure*}

As we did in Fig.~1 in the main text, we ran the seven clustering methods for each number of clusters, $K \in \{2,\cdots , 10\}$, each of the ten sessions, and each of the eight participants. Then, we calculated the $\text{GEV}_{\text{total}}$ and the WCSS. We  show the $\text{GEV}_{\text{total}}$ values in Figs.~\ref{fig:si_gev_compare}(a)--(b) and the WCSS values in Figs.~\ref{fig:si_gev_compare}(c)--(d). Figures~\ref{fig:si_gev_compare}(a) and (c) correspond to the 8-ROI DMN,  and Figs.~\ref{fig:si_gev_compare}(b) and (d) correspond to the 12-ROI DMN. For the 8-ROI DMN, we find that $\text{GEV}_{\text{total}}$ is smaller without global signal removal (shown in Fig.~\ref{fig:si_gev_compare}(a)) than with global signal removal (shown in Figs.~1(a) and (b) in the main text). Moreover, the WCSS is notably larger without global signal removal (shown in Fig.~\ref{fig:si_gev_compare}(c)) than with global signal removal (shown in Figs.~1(d) and (e)).
These results are qualitatively the same for the 12-ROI DMN (see Figs.~\ref{fig:si_gev_compare}(b) and (d), which should be compared with Figs.~1(c) and (f), respectively). Therefore, for the present data, the global signal removal improves the clustering analysis.

\renewcommand\thesection{S3}
\section{Within-participant and between-participant reproducibility of the state-transition dynamics with $K=7$ and $K=10$} \label{si:reproducibility}

\begin{figure*}[htp]
\renewcommand\thefigure{S2}
\centering
\includegraphics[width=1.1\textwidth]{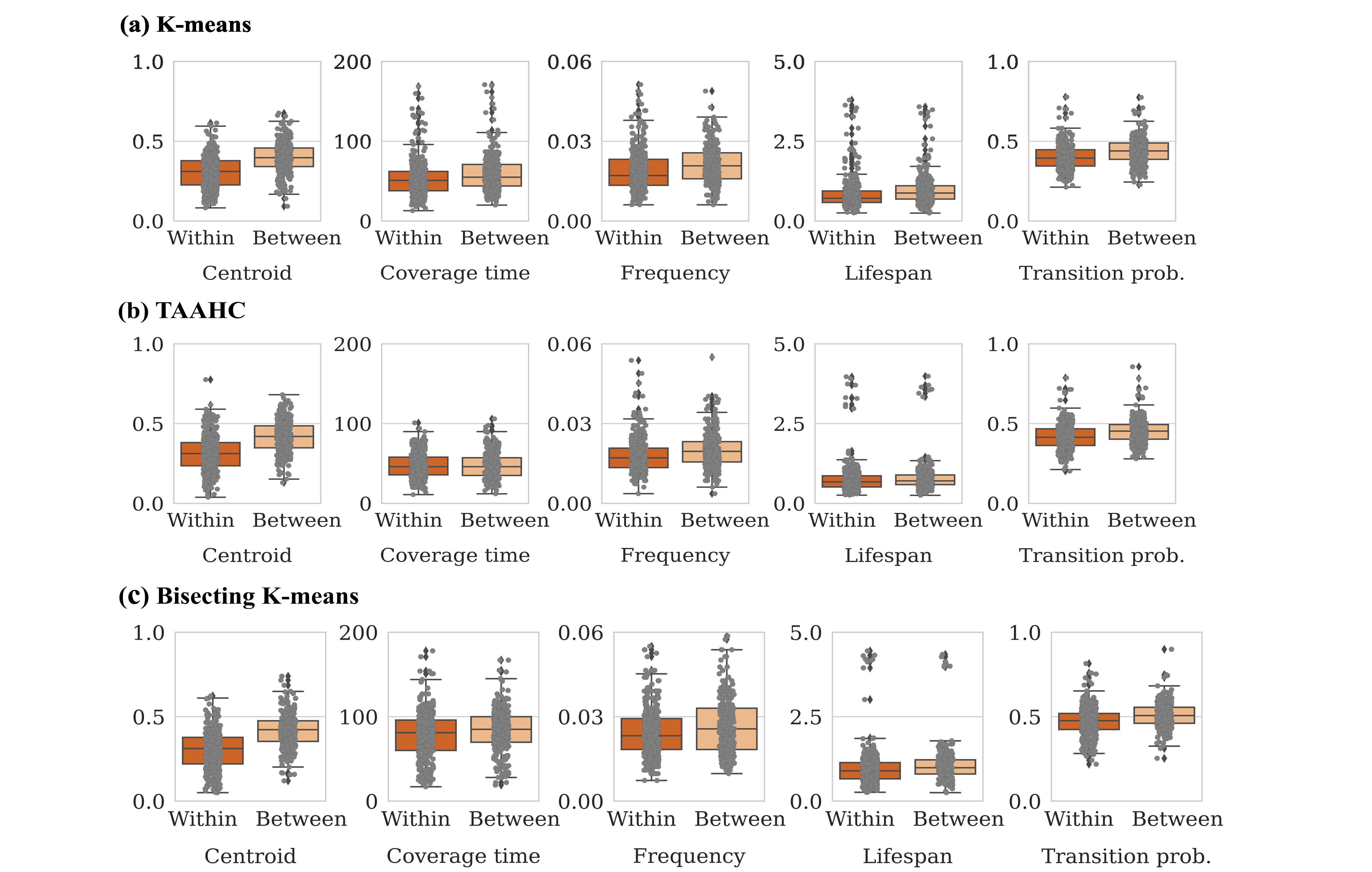}
  \caption{Within-participant and between-participant reproducibility of the state-transition dynamics with $K=7$ states. (a) K-means. (b) TAAHC. (c) Bisecting K-means.
  ``Within'' and ``Between'' indicate the within-participant and between-participant comparisons, respectively. Each box plot shows the minimum, maximum, median, first quartile, and third quartile of the measurements. Each dot represents a session. ``Centroid'' abbreviates the centroid position, and ``Transition prob.'' abbreviates the transition probability matrix.}
  \label{fig:compare-c7}
\end{figure*}

\begin{figure*}[htp]
\renewcommand\thefigure{S3}
\centering
\includegraphics[width=1.1\textwidth]{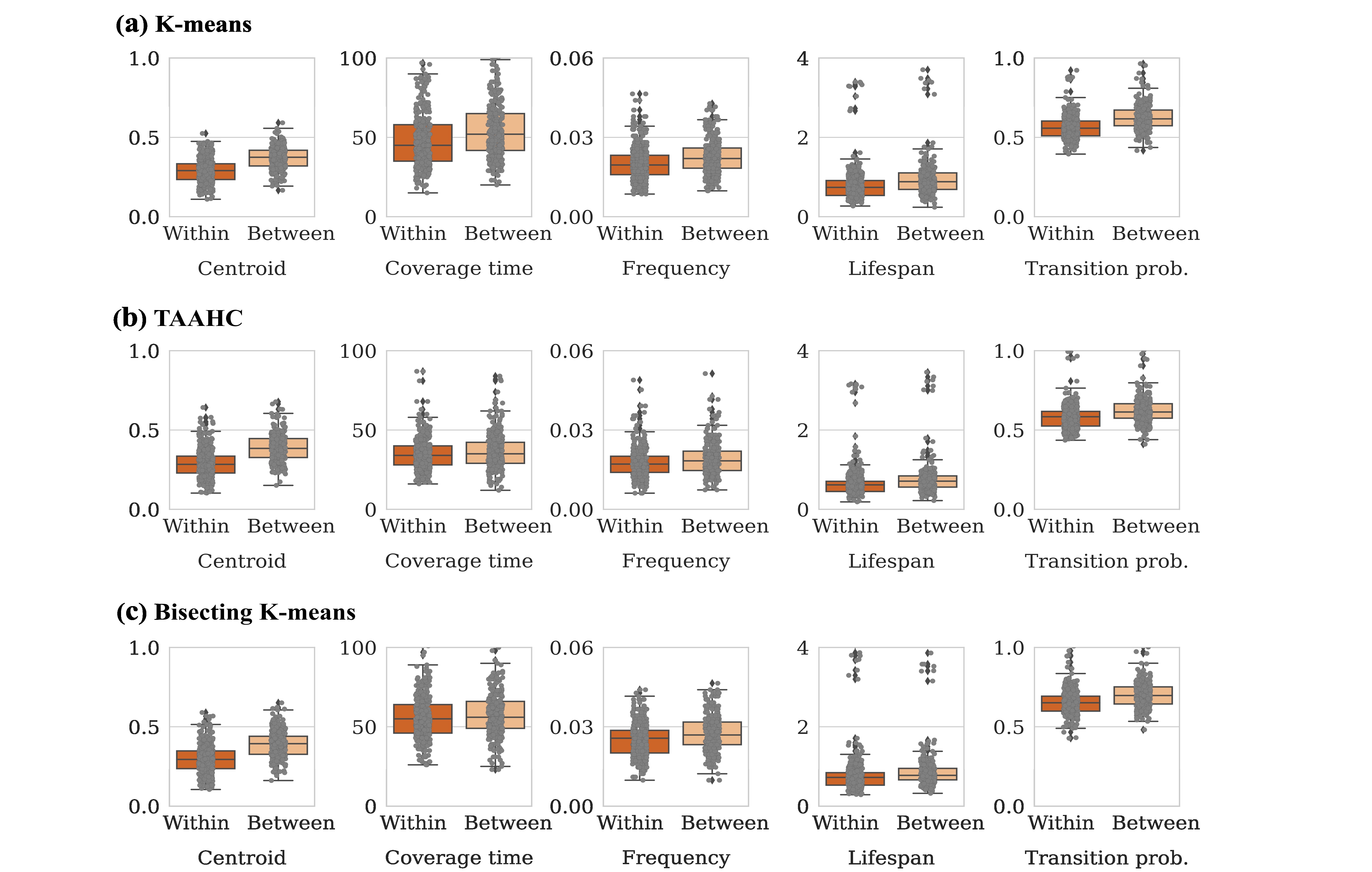}
 \caption{Within-participant and between-participant reproducibility of the state-transition dynamics with $K=10$ states. (a) K-means. (b) TAAHC. (c) Bisecting K-means.}
  \label{fig:compare-c10}
\end{figure*}

We show the distributions of the discrepancy in terms of the five observables for the within-participant
and between-participant comparisons in Figs.~\ref{fig:compare-c7} and \ref{fig:compare-c10} for the number of clusters $K=7$ and $K=10$, respectively.

\renewcommand\thesection{S4}
\section{Robustness test with the Euclidean distance used as dissimilarity between sessions}\label{si:robustness}

We mainly used the dissimilarity based on the $\text{corr}(\bm{c}_\ell, \bm{c}_{\ell'})$ defined by Eq.~(7) 
in the main text to match and compare the set of the centroid positions, $\{ \bm{c}_1, \ldots, \bm{c}_L \}$, between pairs of sessions. In this section, we investigate the robustness of the results of the test-retest reproducibility analysis by running the same analysis with the Euclidean distance between the centroid positions, defined by Eq.~(14) 
 in the main text, to match and compare the set of centroid positions between pairs of sessions. We also use the Euclidean distance to measure the discrepancy in terms of the centroid position between the two optimally matched sessions.

We show the $p$ values for the permutation test for the three clustering methods, five discrepancy measures for the state-transition dynamics, and $K \in \{2, \ldots, 10\}$ in Tables~\ref{tab: msc dmn p-values2}, \ref{tab: msc whole p-values2}, and Table~\ref{tab: hcp whole p-values2} for the DMN extracted from the MSC data, the whole-brain network extracted from the MSC data, and the whole-brain network extracted from the HCP data, respectively. In the tables,  $p <  10^{-4}$ indicates that the deviation of the ND value from $1$ is larger for the original session-to-session comparisons than for all the $10^4$ randomized comparisons. We find that the permutation test results obtained with the Euclidean distance are similar to those obtained with the cosine distance shown in the main text.

Quantitatively, for the DMN extracted from the MSC data, most $p$ values (i.e., 128 out of the 135 comparisons;  94.81\%) were smaller than 0.05 (indicated by *), and 111 of them (i.e., 83.07\% of the 135 comparisons) remained significant after the Bonferroni correction (indicated by ***; equivalent to $p < 0.00037$, uncorrected). Moreover, we obtained $p < 0.001$, indicated with **, in 113 out of the 135 comparisons (i.e., 83.15\%). We also found that the number of significant $p$ values (with $p < 0.05$) for the K-means (45 out of 45) and bisecting K-means (44 out of 45), which differ only by $1$ from each other, was somewhat larger than that for the TAAHC (42 out of 45). After the Bonferroni correction, the number of the significant $p$ values remained larger for the K-means (43 out of 45) than for bisecting K-means (36 out of 45), which is larger than for the TAAHC (32 out of 45).
\begin{table*}
\renewcommand\thetable{S2}
\centering
\caption{Results of the permutation test for the DMN extracted from the MSC data when we used the Euclidean distance to calculate the distance between the centroids of clusters. *: $p<0.05$, uncorrected; **: $p<0.001$, uncorrected; ***: $p<0.00037$, uncorrected (which is equivalent to $p<0.05$, Bonferroni corrected).  We remark that ``Centroid'' and "Trans. prob." abbreviate the centroid's position and the transition probability matrix, respectively.}
\begin{tabular}{>{\centering\arraybackslash}m{0.5cm} >{\centering\arraybackslash}m{0.5cm} >{\centering\arraybackslash}m{1.6cm} >{\centering\arraybackslash}m{1.6cm} >{\centering\arraybackslash}m{1.6cm} >{\centering\arraybackslash}m{1.6cm} >{\centering\arraybackslash}m{1.8cm}}
   \hline
   & $K$ & Centroid & Coverage & Frequency & Lifespan & Trans. prob. \\ 
   \hline                   
\multirow{9}{*}{\begin{turn}{90}K-means\end{turn}} 
				& 2  & $<10^{-4***}$ & $0.0186^*$ & $<10^{-4***}$  &$<10^{-4***}$& $<10^{-4***}$   \\
                                  & 3  & $<10^{-4***}$ & $<10^{-4***}$ & $<10^{-4***}$ & $<10^{-4***}$  & $<10^{-4***}$  \\
                                  & 4  & $<10^{-4***}$ & $0.0014^{*}$  & $<10^{-4***}$         & $<10^{-4***}$   & $<10^{-4***}$    \\
                                  & 5  & $<10^{-4***}$  & $<10^{-4***}$ & $<10^{-4***}$ & $<10^{-4***}$  & $<10^{-4***}$    \\
                                  & 6  & $<10^{-4***}$  & $<10^{-4***}$ & $<10^{-4***}$ & $<10^{-4***}$  & $<10^{-4***}$    \\
                                  & 7  & $<10^{-4***}$  & $<10^{-4***}$ & $<10^{-4***}$ & $<10^{-4***}$  & $<10^{-4***}$    \\
                                  & 8  & $<10^{-4***}$  & $<10^{-4***}$ & $<10^{-4***}$ & $<10^{-4***}$  & $<10^{-4***}$    \\
                                  & 9  & $<10^{-4***}$  & $<10^{-4***}$ & $<10^{-4***}$ & $<10^{-4***}$  & $<10^{-4***}$    \\
                                  & 10 & $<10^{-4***}$  & $<10^{-4***}$ & $<10^{-4***}$ & $<10^{-4***}$  & $<10^{-4***}$    \\
                                  \hline 
\multirow{9}{*}{\begin{turn}{90}TAAHC\end{turn}}  
				& 2  & $<10^{-4***}$     & $0.5528$ & $0.0035^{*}$  & $0.0036^{*}$   & $0.002^{*}$   \\
                                  & 3  & $<10^{-4***}$  & $0.1027$  & $<10^{-4***}$  & $0.0026^{*}$   & $0.0001^{***}$            \\
                                  & 4  & $<10^{-4***}$ & $0.0435^*$         & $<10^{-4***}$         & $<10^{-4***}$ & $<10^{-4***}$    \\
                                  & 5  & $<10^{-4***}$    & $0.0842$         &  $0.0001^{***}$  & $<10^{-4***}$    & $<10^{-4***}$     \\
                                  & 6  & $<10^{-4***}$    & $0.0076^{*} $          & $<10^{-4***}$         & $<10^{-4***}$   & $<10^{-4***}$   \\
                                  & 7  & $<10^{-4***}$    &  0.5548  & $<10^{-4***}$       & $<10^{-4***}$   & $<10^{-4***}$  \\
                                  & 8  & $<10^{-4***}$      &  $0.0718$      & $<10^{-4***}$       & $<10^{-4***}$      & $<10^{-4***}$  \\
                                  & 9  & $<10^{-4***}$     & $0.0434^*$        &  $<10^{-4***}$        & $<10^{-4***}$  & $<10^{-4***}$  \\
                                  & 10 & $<10^{-4***}$  & $0.2352$     & $<10^{-4***}$ & $<10^{-4***}$  &  $<10^{-4***}$       \\
                                  \hline 
\multirow{9}{*}{\begin{turn}{90}Bisecting K-means\end{turn}} 
				& 2 & $<10^{-4***}$  &  $0.0287^*$ &  $<10^{-4***}$   & $<10^{-4***}$& $<10^{-4***}$   \\
                                  & 3  & $<10^{-4***}$   & $0.0001^{***}$ & $<10^{-4***}$  & $<10^{-4***}$ & $<10^{-4***}$               \\
                                  & 4  & $<10^{-4***}$   &  0.0584   &$<10^{-4***}$   & $<10^{-4***}$     & $<10^{-4***}$       \\
                                  & 5  & $<10^{-4***}$  & $0.0419^{*}$  & $0.0003^{***}$ & $<10^{-4***}$   & $<10^{-4***}$    \\
                                  & 6  & $<10^{-4***}$  & $0.018^{*}$   & $0.0001^{***}$ & $0.0004^{**}$  & $<10^{-4***}$      \\
                                  & 7  & $<10^{-4***}$   & $0.0005^{**}$         & $<10^{-4***}$        & $<10^{-4***}$     & $<10^{-4***}$ \\
                                  & 8  & $<10^{-4***}$     & $0.0037^{*}$   & $0.0002^{***}$     & $0.0001^{***}$      & $<10^{-4***}$    \\
                                  & 9  & $<10^{-4***}$    & $0.0072^{*}$  &  $0.0001^{***}$    & $<10^{-4***}$   & $<10^{-4***}$  \\
                                  & 10 & $<10^{-4***}$    & $0.0067^{*}$          & $<10^{-4***}$    & $<10^{-4***}$    & $<10^{-4***}$   \\
                                  \hline      
\end{tabular}
\label{tab: msc dmn p-values2}
 \end{table*}
 
\begin{table*}
\renewcommand\thetable{S3}
\centering
\caption{Results of the permutation test for the whole-brain network extracted from the MSC data when we used the Euclidean distance to calculate the distance between the centroids of clusters. *: $p<0.05$, uncorrected; **: $p<0.001$, uncorrected; ***: $p<0.00037$, uncorrected (which is equivalent to $p<0.05$, Bonferroni corrected).  We remark that ``Centroid'' and "Trans. prob." abbreviate the centroid's position and the transition probability matrix, respectively.}
\begin{tabular}{>{\centering\arraybackslash}m{0.5cm} >{\centering\arraybackslash}m{0.5cm} >{\centering\arraybackslash}m{1.6cm} >{\centering\arraybackslash}m{1.6cm} >{\centering\arraybackslash}m{1.6cm} >{\centering\arraybackslash}m{1.6cm} >{\centering\arraybackslash}m{1.8cm}}
   \hline
   & $K$ & Centroid & Coverage & Frequency & Lifespan & Trans. prob. \\ 
   \hline 
\multirow{9}{*}{\begin{turn}{90}K-means\end{turn}}  
	  & 2 & $<10^{-4***}$ & $0.0368^*$ & $<10^{-4***}$ & $<10^{-4***}$ & $<10^{-4***}$ \\ 
        ~ & 3 & $<10^{-4***}$ &$ 0.0402^*$ & $<10^{-4***}$ & $<10^{-4***}$ & $<10^{-4***}$ \\ 
        ~ & 4 & $<10^{-4***}$ & $0.0353^*$ & $<10^{-4***}$ & $<10^{-4***}$ & $<10^{-4***}$ \\ 
        ~ & 5 & $<10^{-4***}$  & $<10^{-4***} $& $<10^{-4***}$ & $<10^{-4***}$ & $<10^{-4***}$ \\ 
        ~ & 6 & $<10^{-4***}$  & $0.0007^{**} $& $<10^{-4***}$ & $<10^{-4***}$ & $<10^{-4***}$ \\ 
        ~ & 7 & $<10^{-4***}$ & $0.0016^{*}$ & $<10^{-4***}$ & $<10^{-4***}$ & $<10^{-4***}$ \\ 
        ~ & 8 &  $<10^{-4***}$  & $0.0004^{**} $& $<10^{-4***}$ & $<10^{-4***}$ & $<10^{-4***}$ \\ 
        ~ & 9 &  $<10^{-4***}$  & $10^{-4***} $& $<10^{-4***}$ & $<10^{-4***}$ & $<10^{-4***}$ \\ 
        ~ & 10 &  $<10^{-4***}$  & $0.0003^{***} $& $<10^{-4***}$ & $<10^{-4***}$ & $<10^{-4***}$ \\  \hline 
\multirow{9}{*}{\begin{turn}{90}TAAHC\end{turn}}  
	  & 2 & $<10^{-4***}$  & $0.0776$ & $0.0001^{***}$ & $<10^{-4***}$ & $0.0001^{***}$ \\ 
        ~ & 3 & $<10^{-4***}$ & $0.0463^*$ & $<10^{-4***}$ & $0.0003^{***}$ & $<10^{-4***}$ \\ 
        ~ & 4 & $<10^{-4***}$ & 0.0940 & $<10^{-4***}$ & $0.0011^{*}$ & $0.0004^{**}$ \\ 
        ~ & 5 & $<10^{-4***}$  & $0.0813$ & $<10^{-4***}$ & $<10^{-4***}$ & $<10^{-4***}$\\ 
        ~ & 6 & $<10^{-4***}$ & $0.7868$ & $<10^{-4***}$ & $0.0011^{*}$ & $<10^{-4***}$ \\ 
        ~ & 7 & $<10^{-4***}$ & $0.8686$ & $<10^{-4***}$ & $<10^{-4***}$ & $<10^{-4***}$ \\ 
        ~ & 8 & $<10^{-4***}$ & 0.0875 & $<10^{-4***}$ & $<10^{-4***}$ & $<10^{-4***}$ \\ 
        ~ & 9 & $<10^{-4***}$  & $0.0403^*$& $<10^{-4***}$ & $<10^{-4***}$ & $<10^{-4***}$ \\ 
        ~ & 10 & $<10^{-4***}$ & $0.036^*$ & $<10^{-4***}$ & $<10^{-4***}$ & $<10^{-4***}$ \\  \hline 
\multirow{9}{*}{\begin{turn}{90}Bisecting K-means\end{turn}} 
	& 2 & $<10^{-4***}$ &$0.0303^*$ & $<10^{-4***}$ & $<10^{-4***}$ & $<10^{-4***}$ \\ 
        ~ & 3 & $<10^{-4***}$  & $0.2482$ & $<10^{-4***}$ & $0.0041^{*}$ & $0.0174^{*}$ \\ 
        ~ & 4 & $<10^{-4***}$  & $0.0576$ & $<10^{-4***}$ & $<10^{-4***}$ & $<10^{-4***}$ \\ 
        ~ & 5 & $<10^{-4***}$ & 0.2599 & $0.0002^{***}$ & $0.0003^{***}$ & $0.0483^*$ \\ 
        ~ & 6 & $<10^{-4***}$ & $0.0658$ & $<10^{-4***}$ & $<10^{-4***}$& $<10^{-4***}$ \\ 
        ~ & 7 & $<10^{-4***}$  & $0.6427$ & $0.0002^{***}$ & $<10^{-4***}$ & $0.0001^{***}$ \\ 
        ~ & 8 & $<10^{-4***}$  & $<10^{-4***}$ & $<10^{-4***}$ & $<10^{-4***}$ & $<10^{-4***}$ \\ 
        ~ & 9 & $<10^{-4***}$  & 0.0905 & $0.0018^{*}$  & $<10^{-4***}$& $<10^{-4***}$ \\ 
        ~ & 10 & $<10^{-4***}$ & $0.0031^{*}$ & $<10^{-4***}$ & $<10^{-4***}$ & $<10^{-4***}$ \\   \hline     
\end{tabular}
\label{tab: msc whole p-values2}
 \end{table*}

\begin{table*}
\renewcommand\thetable{S4}
\centering
\caption{Results of the permutation test for the whole-brain network extracted from the HCP data when we used the Euclidean distance to calculate the distance between the centroids of clusters. T*: $p<0.05$, uncorrected; **: $p<0.001$, uncorrected; ***: $p<0.00037$, uncorrected (which is equivalent to $p<0.05$, Bonferroni corrected).  We remark that ``Centroid'' and "Trans. prob." abbreviate the centroid's position and the transition probability matrix, respectively.}
\begin{tabular}{>{\centering\arraybackslash}m{0.5cm} >{\centering\arraybackslash}m{0.5cm} >{\centering\arraybackslash}m{1.6cm} >{\centering\arraybackslash}m{1.6cm} >{\centering\arraybackslash}m{1.6cm} >{\centering\arraybackslash}m{1.6cm} >{\centering\arraybackslash}m{1.8cm}}
   \hline
   & $K$ & Centroid & Coverage & Frequency & Lifespan & Trans. prob. \\ 
   \hline  
\multirow{9}{*}{\begin{turn}{90}K-means\end{turn}} 
	   & 2 & $<10^{-4***}$ & $0.0023^{*}$ & $<10^{-4***}$ & $<10^{-4***}$ & $<10^{-4***}$ \\
        ~ & 3 & $<10^{-4***}$ &$ < 10^{-4***}$ & $<10^{-4***}$ & $<10^{-4***}$ & $<10^{-4***}$ \\ 
        ~ & 4 & $<10^{-4***}$  & $<10^{-4***}$ & $<10^{-4***}$ & $<10^{-4***}$ & $<10^{-4***}$ \\ 
        ~ & 5 & $<10^{-4***}$ & $<10^{-4***}$ & $<10^{-4***}$ & $<10^{-4***}$ & $<10^{-4***}$ \\ 
        ~ & 6 & $<10^{-4***}$  & $<10^{-4***}$ & $<10^{-4***}$ & $<10^{-4***}$ & $<10^{-4***}$ \\ 
        ~ & 7 & $<10^{-4***}$  & $<10^{-4***}$ & $<10^{-4***}$ & $<10^{-4***}$ & $<10^{-4***}$ \\ 
        ~ & 8 & $<10^{-4***}$  & $<10^{-4***}$ & $<10^{-4***}$ & $<10^{-4***}$ & $<10^{-4***}$ \\ 
        ~ & 9 & $<10^{-4***}$ & $<10^{-4***}$ & $<10^{-4***}$ & $<10^{-4***}$ & $<10^{-4***}$ \\ 
        ~ & 10 & $<10^{-4***}$ & $<10^{-4***}$ & $<10^{-4***}$ & $<10^{-4***}$ & $<10^{-4***}$ \\  \hline 
\multirow{9}{*}{\begin{turn}{90}TAAHC\end{turn}}  
   	  & 2 & $<10^{-4***}$ & $<10^{-4***}$ & $<10^{-4***}$ & $<10^{-4***}$ & $<10^{-4***}$ \\ 
        ~ & 3 & $<10^{-4***}$  & 0.0952 & $<10^{-4***}$ & $<10^{-4***}$ & $<10^{-4***}$ \\ 
        ~ & 4 & $<10^{-4***}$  & $0.0159^{*}$ & $<10^{-4***}$ & $<10^{-4***}$ & $<10^{-4***}$ \\ 
        ~ & 5 & $<10^{-4***}$  & $0.1318$ & $<10^{-4***}$ & $<10^{-4***}$& $<10^{-4***}$ \\ 
        ~ & 6 & $<10^{-4***}$  & $0.0191^*$ & $<10^{-4***}$ & $<10^{-4***}$ & $<10^{-4***}$ \\ 
        ~ & 7 & $<10^{-4***}$  & $0.0096^{*}$ & $<10^{-4***}$ & $<10^{-4***}$ & $<10^{-4***}$ \\ 
        ~ & 8 & $<10^{-4***}$  & $0.0007^{**}$ & $<10^{-4***}$ & $<10^{-4***}$ & $<10^{-4***}$ \\ 
        ~ & 9 & $<10^{-4***}$ & $0.0003^{***}$ & $<10^{-4***}$ & $<10^{-4***}$ & $<10^{-4***}$ \\ 
        ~ & 10 &  $<10^{-4***}$ & $0.0008^{**}$ & $<10^{-4***}$ & $<10^{-4***}$ & $<10^{-4***}$ \\ \hline 
\multirow{9}{*}{\begin{turn}{90}Bisecting K-means\end{turn}} 
	   & 2 & $<10^{-4***}$  & $0.0016^{*}$ & $<10^{-4***}$ & $<10^{-4***}$ & $<10^{-4***}$ \\ 
        ~ & 3 & $<10^{-4***}$  & $0.013^*$ & $<10^{-4***}$ & $<10^{-4***}$ & $<10^{-4***}$ \\ 
        ~ & 4 & $<10^{-4***}$ &  $<10^{-4***}$ & $<10^{-4***}$ & $<10^{-4***}$ & $<10^{-4***}$ \\ 
        ~ & 5 & $<10^{-4***}$  & $<10^{-4***}$ & $<10^{-4***}$ & $<10^{-4***}$ & $<10^{-4***}$ \\ 
        ~ & 6 & $<10^{-4***}$  & $0.0001^{***}$ & $<10^{-4***}$ & $<10^{-4***}$ & $<10^{-4***}$ \\ 
        ~ & 7 & $<10^{-4***}$  & $0.0118^{*}$ & $<10^{-4***}$ & $<10^{-4***}$ & $<10^{-4***}$ \\ 
        ~ & 8 & $<10^{-4***}$ & $<10^{-4***}$ & $<10^{-4***}$& $<10^{-4***}$ & $<10^{-4***}$ \\ 
        ~ & 9 &  $<10^{-4***}$ & $<10^{-4***}$ & $<10^{-4***}$ & $<10^{-4***}$ & $<10^{-4***}$ \\ 
        ~ & 10 & $<10^{-4***}$ & $<10^{-4***}$ & $<10^{-4***}$ & $<10^{-4***}$ & $<10^{-4***}$ \\\hline     
\end{tabular}
\label{tab: hcp whole p-values2}
 \end{table*}
 
 \renewcommand\thesection{S5}
 \section{Impact of the number of ROIs in the DMN on test-retest reproducibility}\label{si:impact_dr}
In this section, we explore the influence of the number of ROIs given by different coordinate systems on test-retest reproducibility. To this end, we use the DMN with 59 ROIs extracted from MSC dataset that is part of the 264-ROI system that we have used for constructing the whole-brain network~\cite{power2011functional}. We followed the same analysis pipeline employed in the main text. For the 59-ROI DMN, we solely removed the global signal without averaging the time series of each left- and right-hemispheric pair of symmetric ROIs.

We show the $p$ values for the permutation test across three clustering methods, five discrepancy measures for the state-transition dynamics, and $K \in\{2, \cdots , 10\}$ in Table~\ref{tab: msc 59dmn p-values}. We find the within-participant reproducibility relative to the between-participant reproducibility is similarly high with the present 59-ROI DMN compared to the case of the 12-ROI DMN (see Table 2 in the main text), except that the coverage in the case of K-means yields poor results for the 59-ROI DMN but not for the 12-ROI DMN.

\begin{table*}
\renewcommand\thetable{S5}
\centering
\caption{Results of the permutation test for the 59-ROI DMN extracted from the MSC data. *: $p<0.05$, uncorrected; **: $p<0.001$, uncorrected; ***: $p<0.00037$ uncorrected (which is equivalent to $p<0.05$, Bonferroni corrected). We remark that ``Centroid'' and "Trans. prob." abbreviate the centroid's position and the transition probability matrix, respectively.}
\begin{tabular}{>{\centering\arraybackslash}m{0.5cm} >{\centering\arraybackslash}m{0.5cm} >{\centering\arraybackslash}m{1.6cm} >{\centering\arraybackslash}m{1.6cm} >{\centering\arraybackslash}m{1.6cm} >{\centering\arraybackslash}m{1.6cm} >{\centering\arraybackslash}m{1.8cm}}
   \hline
   & $K$ & Centroid & Coverage & Frequency & Lifespan & Trans. prob. \\ 
   \hline  
   \multirow{9}{*}{\begin{turn}{90}K-means\end{turn}} 
				& 2  & $<10^{-4***}$ & $0.0285^*$ & $<10^{-4***}$  &$0.0002^{***}$& $<10^{-4***}$   \\
                                  & 3  & $<10^{-4***}$ & $0.5994$ & $<10^{-4***}$ & $<10^{-4***}$  & $<10^{-4***}$  \\
                                  & 4  & $<10^{-4***}$ & $0.5626$  & $<10^{-4***}$         & $0.0001^{***}$   & $<10^{-4***}$    \\
                                  & 5  & $<10^{-4***}$  & $0.7903$ & $<10^{-4***}$ & $<10^{-4***}$  & $<10^{-4***}$    \\
                                  & 6  & $<10^{-4***}$  & $0.0004^{**}$ & $<10^{-4***}$ & $<10^{-4***}$  & $<10^{-4***}$    \\
                                  & 7  & $<10^{-4***}$  & $0.0232^*$ & $<10^{-4***}$ & $<10^{-4***}$  & $<10^{-4***}$    \\
                                  & 8  & $<10^{-4***}$  & $0.1162$ & $<10^{-4***}$ & $<10^{-4***}$  & $<10^{-4***}$    \\
                                  & 9  & $<10^{-4***}$  & $0.2576$ & $<10^{-4***}$ & $<10^{-4***}$  & $<10^{-4***}$    \\
                                  & 10 & $<10^{-4***}$  & $0.2229$ & $<10^{-4***}$ & $<10^{-4***}$  & $<10^{-4***}$    \\
                                  \hline 
\multirow{9}{*}{\begin{turn}{90}TAAHC\end{turn}}  
				& 2  & $<10^{-4***}$     & $0.3201$ &$<10^{-4***}$  & $<10^{-4***}$  & $<10^{-4***}$   \\
                                  & 3  & $<10^{-4***}$  & $0.4577$  & $<10^{-4***}$  & $0.0001^{***}$   & $<10^{-4***}$ \\
                                  & 4  & $<10^{-4***}$ & $0.1850$         & $<10^{-4***}$         & $<10^{-4***}$ & $<10^{-4***}$    \\
                                  & 5  & $<10^{-4***}$    & $0.3747$         &  $<10^{-4***}$ & $<10^{-4***}$    & $<10^{-4***}$     \\
                                  & 6  & $<10^{-4***}$    & $0.4879 $          & $<10^{-4***}$         & $<10^{-4***}$   & $<10^{-4***}$   \\
                                  & 7  & $<10^{-4***}$    &  0.3990  & $<10^{-4***}$       & $<10^{-4***}$   & $<10^{-4***}$  \\
                                  & 8  & $<10^{-4***}$      &  $0.5510$      & $<10^{-4***}$       & $<10^{-4***}$      & $<10^{-4***}$  \\
                                  & 9  & $<10^{-4***}$     & $0.6263$        &  $<10^{-4***}$        & $<10^{-4***}$  & $<10^{-4***}$  \\
                                  & 10 & $<10^{-4***}$  & $0.7597$     & $<10^{-4***}$ & $<10^{-4***}$  &  $<10^{-4***}$       \\
                                  \hline 
\multirow{9}{*}{\begin{turn}{90}Bisecting K-means\end{turn}} 
				& 2 & $<10^{-4***}$  &  $0.0184^*$ &  $<10^{-4***}$   & $<10^{-4***}$& $<10^{-4***}$   \\
                                  & 3  & $<10^{-4***}$   & $0.0311^{*}$ & $<10^{-4***}$  & $<10^{-4***}$ & $<10^{-4***}$               \\
                                  & 4  & $<10^{-4***}$   &  0.7056   &$<10^{-4***}$   & $0.0002^{***}$     & $<10^{-4***}$       \\
                                  & 5  & $<10^{-4***}$  & $0.1006$  & $<10^{-4***}$ & $<10^{-4***}$   & $<10^{-4***}$    \\
                                  & 6  & $<10^{-4***}$  & $0.5694$   & $<10^{-4***}$ & $0.0024^{*}$  & $<10^{-4***}$      \\
                                  & 7  & $<10^{-4***}$   & $0.0343^{*}$         & $<10^{-4***}$        & $<10^{-4***}$     & $<10^{-4***}$ \\
                                  & 8  & $<10^{-4***}$     & $0.0006^{**}$   & $<10^{-4***}$     &$<10^{-4***}$    & $<10^{-4***}$    \\
                                  & 9  & $<10^{-4***}$    & $0.0053^{*}$  &  $<10^{-4***}$   & $0.0001^{***}$   & $<10^{-4***}$  \\
                                  & 10 & $<10^{-4***}$    & $0.2402$          & $0.0001^{***}$    & $0.0006^{**}$    & $<10^{-4***}$   \\
                                  \hline     
\end{tabular}
\label{tab: msc 59dmn p-values}
 \end{table*}

 \end{multicols}
%

\end{document}